\documentclass[onecolumn,showpacs,preprintnumbers,amsmath,amssymb,superscriptaddress]{revtex4}
\usepackage{graphicx}
\usepackage{dcolumn}
\usepackage{bm}
\usepackage{longtable}

\newcommand{\beq}{\begin{equation}}
\newcommand{\eeq}{\end{equation}}
\newcommand{\bea}{\begin{eqnarray}}
\newcommand{\eea}{\end{eqnarray}}
\def\be{\begin{equation}}
\def\ee{\end{equation}}

\begin{document}

\preprint{APS/123-QED}

\title{Parallel processing in immune networks}

\author{Elena Agliari}
\affiliation{Dipartimento di Fisica, Universit\`{a} degli Studi di Parma, viale G. Usberti 7, 43100 Parma, Italy}
\affiliation{INFN, Gruppo Collegato di Parma, viale G. Usberti 7, 43100 Parma, Italy}
\author{Adriano Barra}
\affiliation{Dipartimento di Fisica, Sapienza Universit\`{a} di Roma, Piazzale Aldo Moro 2, 00185, Roma, Italy}
\author{Silvia Bartolucci}
\affiliation{Mathematics Department, King's College London, Strand, London $WC2R$ \  $2LS$, U.K.}
\affiliation{Institute for Mathematical and Molecular Biomedicine, King's College London, Guy's Campus
London $SE1$ \  $1UL$, U.K.}
\author{Andrea Galluzzi}
\affiliation{Dipartimento di Fisica, Sapienza Universit\`{a} di Roma, Piazzale Aldo Moro 2, 00185, Roma, Italy}
\author{Francesco Guerra}
\affiliation{Dipartimento di Fisica, Sapienza Universit\`{a} di Roma, Piazzale Aldo Moro 2, 00185, Roma, Italy}
\author{Francesco Moauro}
\affiliation{Dipartimento di Fisica, Sapienza Universit\`{a} di Roma, Piazzale Aldo Moro 2, 00185, Roma, Italy}

\date{\today}

\begin{abstract}
In this work we adopt a statistical mechanics approach to investigate basic, systemic features exhibited by adaptive immune systems. The lymphocyte network made by B-cells and T-cells is modeled by a bipartite spin-glass, where, following biological prescriptions, links connecting B-cells and T-cells are sparse. Interestingly, the dilution performed on links is shown to make the system able to orchestrate parallel strategies to fight several pathogens at the same time; this multitasking capability constitutes a remarkable, key property of immune systems as multiple antigens are always present within the host.
We also define the stochastic process ruling the temporal evolution of lymphocyte activity, and show its relaxation toward an equilibrium measure allowing statistical mechanics investigations. Analytical results are compared with Monte Carlo simulations and signal-to-noise outcomes showing overall excellent agreement.
Finally, within our model, a rationale for the experimentally well-evidenced correlation between lymphocytosis and autoimmunity is achieved; this sheds further light on the systemic features exhibited by immune networks.
\end{abstract}



 \maketitle

\section{Introduction}

While the first half of the XIX century saw triumphal discoveries in physics, ranging from quantum mechanics and general relativity up to the discovery of chaos, the second half has probably been dragged by biology: Among its several fields of investigation, immunology (both theoretical and experimental) is currently one of the most promising.

The immune system constitutes the defensive army of host against pathogens as bacteria, virus, fungi or deranged cells. In higher organisms one usually distinguishes between the innate immune system and the adaptive immune system; the latter is able to mount a specific response against diverse and evolving pathogens. The adaptive immune system is basically a network of lymphocytes exchanging chemical signals and proteins such as cytokines or antibodies. In particular,  B lymphocytes produce antibodies
and are grouped into clones: all cells belonging to the same clone produce the same antibody, while different clones produce different antibodies. When a pathogen enters the body, the best-matching clone expands and its cells secrete the antibody able to chemically bind the pathogen, hence (possibly) avoiding the propagation of the infection. If the pathogen has already infected a host cell, the latter is killed (e.g. via lysis) by Killer lymphocytes and order is restored. B cells and Killer cells make up the so called ``effector branches'', whose activation can take place only if another signal (beyond the presence of the pathogen) occurs. This signal is prompted by another subset of lymphocytes, i.e. helper T lymphocytes, which, devoid of any cytotoxic or phagocytic activity, can coordinate/modulate the immune response by exchanging  with the effector branches either eliciting (e.g. interleukin-$4$ cytokine) or suppressive (e.g. interleukin-$10$ cytokine) messages \cite{abbas}.

\begin{figure}[tb]\label{silvia}
\begin{center}
\includegraphics[width=9cm]{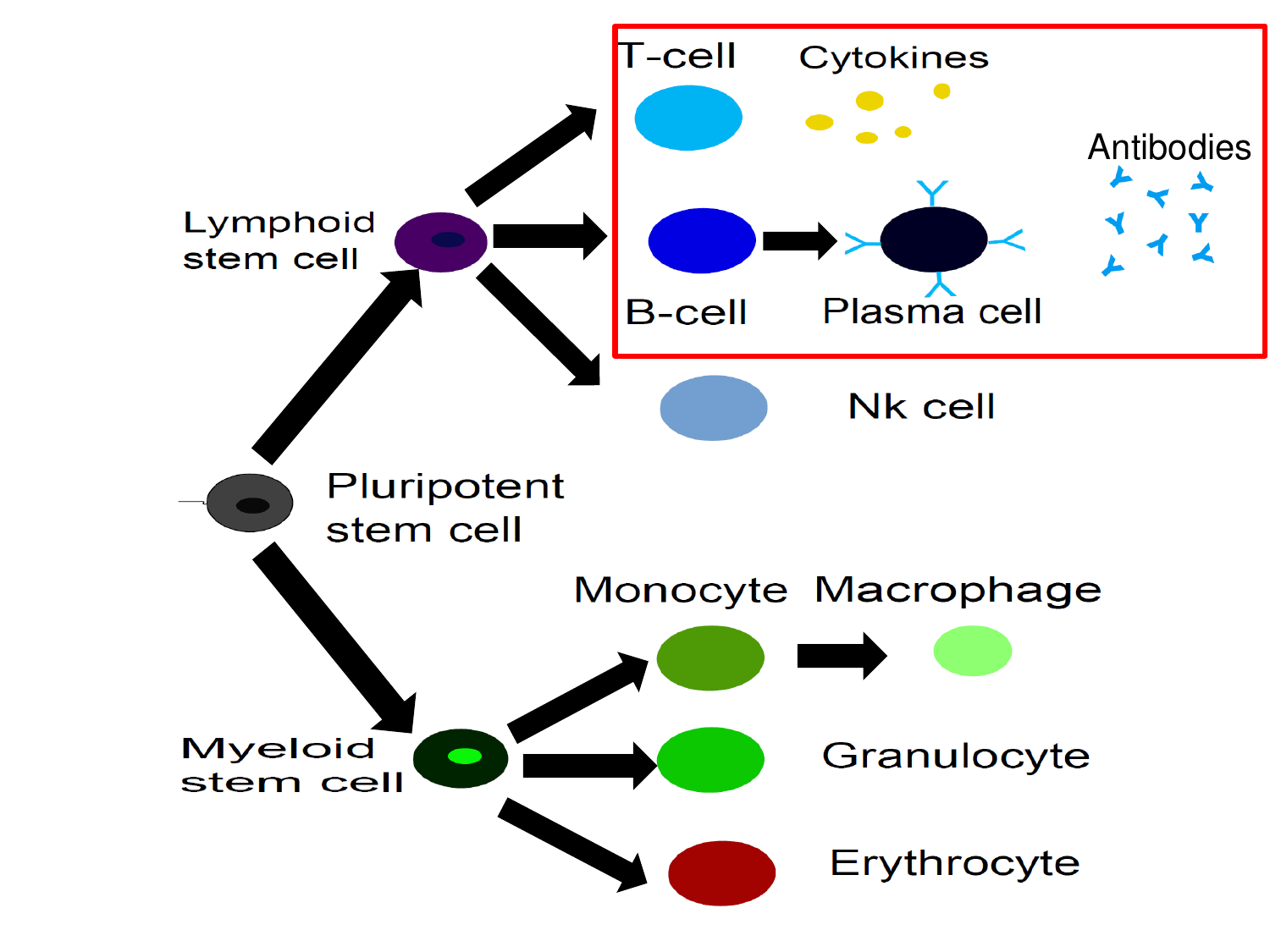}
\caption{(Color on line) Main constituents of the immune system: From the stem cell (left) two branches develop, roughly speaking defining the primary immune response (lower branch) and the secondary adaptive one (higher branch). In the box there are B and T cells, whose properties are the subject of the present investigation.}
\end{center}
\end{figure}

Given the large amount of its constituents (e.g. the complete B-repertoire in humans is estimated to range in $10^{8}-10^{10}$ clones) and the interest in understanding global ``collective" features of the immune system thought of ``as a whole", scientists are becoming attracted towards the potentiality of statistical-mechanics approaches even in this area of theoretical biology (see e.g. \cite{noi1,Agliari-PLOS2013,kardar,bialek,parisi}).
%
%
Accordingly, here we deepen and extend the model introduced in \cite{JTB}, which is focused on the adaptive immune response performed by B-cells and helpers interacting via antibodies and cytokines.
Clearly, such a model is far from being a complete representation of the whole immune system (which is built on a huge number of different constituents \cite{abbas}, see Fig. \ref{silvia}), yet it is able to capture, as emergent features, some collective properties of real systems.

From a mathematical perspective, the model is based on bipartite spin-glasses \cite{MPV} and on their equivalence with information processing systems \cite{peter} as Boltzmann machines and Neural networks.
As we will explain, the interactions between helper cells and B-cells via cytokines give rise to an effective Hebbian structure among helpers alone, where the latter, under proper conditions, relax towards a ``retrieval state'', meant as the proper activation of a single B-clone \cite{JTB}. Interestingly, bypassing the mean-field approximation, where each helper interacts  with each B-clone, toward a description where each helper interacts only with a fraction of the available repertoire of B-clones (as biologically required), makes the helpers able to activate multiple clones to fight several pathogens simultaneously. This means that helpers can perform parallel retrievals (i.e. strategies, instructions for the B-cells) at the same time, without falling into spurious states (i.e. errors), typical of the underlying glassy nature of neural networks.
This is the first network able to accomplish this task  as an emergent property, and its study could contribute toward a rationale of the systemic properties of immune networks.

Such a theoretical framework aims to mirror Medzhitov research which, ultimately, seeks to understand the rules of engagement when our body simultaneously deals with multiple infections in vivo (see e.g. \cite{med1,med2}).
\newline
Indeed, despite a clear interest in compassing these multitasking capabilities of the immune system, and despite dramatic evidences of their failures (e.g. in advanced H.I.V. progression several opportunistic infections, altogether, become fatal \cite{silvia1,silvia2,silvia3}), much efforts are still required, from both experimental and theoretical viewpoints, in order to get a clear picture. In this work we will trying to tackle the problem via a statistical mechanics approach, focusing on multitasking associative networks.

As a necessary first step, here we will restrict to the ``low storage regime'',
where the number of helpers is considerably larger than that of B-cells. This is clearly an approximation, however it still poses the basis for a general comprehension of parallel processing in biological devices and opens interesting questions on the capabilities of diluted (real) networks.

The paper is organized as follows. In section $2$ we review the minimal, fully-connected model previously introduced in \cite{JTB}, while in Sec.~$3$ we explain how dilution is introduced and we scaffold the statistical-mechanics analysis; then, in Sec.~$4$ we study in details the parallel retrieval performed by the system and in Sec.~$5$ we give some insights in the numerical methods exploited; in Sec.~$6$, we model the occurrence of lymphocytosis and see how it might be related autoimmune phenomena; finally, Sec.~$7$ is left for discussions on results and on future perspectives. Technical details concerning the analytical and numerical solutions of the model are collected in the appendices.

\section{The minimal model: features and limitations}

In this section we briefly review a minimal model for the response of the adaptive immune system \cite{JTB}. Since in the original model there is full symmetry between the effector branches made by B cells and killer cells, respectively, in the following we consider only the B-branch, so that the protagonists of the paper will be B cells, helper cells and their chemical messengers.

\begin{figure}[tb]\label{Uovo}
\begin{center}
{\includegraphics[width=16cm]{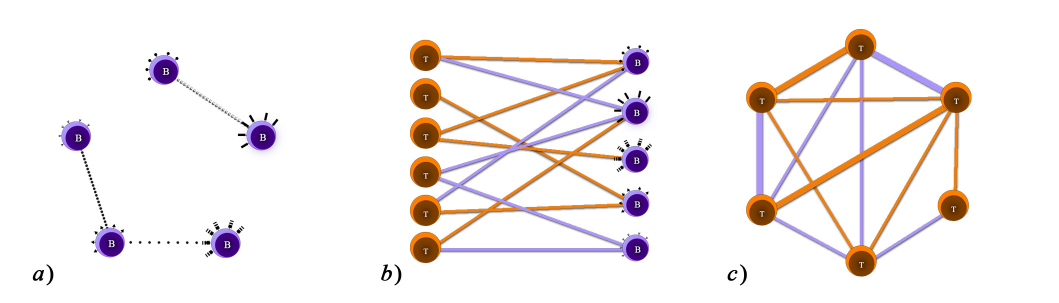}}
\caption{(Color on line) Example of networks with $B=5$ and $H=6$. Panel $a$: Network for B-cells pairwise connected through idiotypic imitative couplings (dotted lines); notice that the network is underpercolated and made of two components. Each node is associated to a different specificity encoded by the cell receptor. Panel $b$: Bipartite network for B-cells and T-cells interacting through excitatory (brighter link) or inhibitory (darker link) signals $\xi$. For instance, assuming that all T cells are active, i.e. $h_i=1$, the B clone represented by the upmost node on the right receives two inhibitory signals and one excitatory signal.
Notice that, due to dilution, only a fraction of the all possible $B \times H$ links are present. Panel $c$: Monopartite network for T-cells obtained by the bipartite graph in panel $b$ through Eq.~\ref{eq:hebb1}, which directly relates the pattern of interactions $\xi$ between helpers and B cells to the couplings $\mathbf{J}$ between helpers. In the monopartite, weighted graph, T cells are pairwise connected through imitative (i.e., $J_{ij}>0$, brighter link) or anti-imitative (i.e., $J_{ij}<0$, darker link) couplings $J$, whose magnitude is rendered by the thickness of the link.}
\end{center}
\end{figure}

There are $B$ different B-clones  and each of them is built of by an amount of identical B-cells.
We call $b_{0,\mu}$ the size of the $\mu^{th}$ clone in normal conditions, namely the background reference value, while the size measured at any arbitrary state is referred to as $b_{\mu}$, in such a way that the difference $(b_{\mu}-b_{0,\mu})$ can be either positive (if the clone has expanded) or negative (if the clone has shrunk). Since the range of reference values is much smaller than the extent of variation of each clone size, it is possible to assume that the normal size of each clone is the same, namely $b_{0,\mu} = b_{0}, \, \forall \mu$; moreover, in an healthy state, we can set $b_0$ equal to zero without loss of generality. In general, $(b_{\mu}-b_{0})$ is a real variable,
ranging in $(b_{\mu}-b_{0}) \in (-\infty, \infty)$, and it will be looked at as a ``soft spin''.
\newline
B lymphocytes interact with each other through an effective, ``imitative'' coupling $\mathbf{I}$, with elements $I_{\mu,\nu}$, such that when a clone $\mu$ has expanded and a large amount of related antibodies is secreted, these can act upon the other clones $\nu$ and prompt their expansion \cite{noi1,antonio, PRE}. The resulting network of interactions is often referred to as idiotypic network (see Fig.~\ref{Uovo}, panel $a$).

As for helper cells, we analogously introduce a set of quantities representing the status of each clone: we call $h_i$ the activity of the $i^{th}$ helper clone. However, differently from B cells, helper cells interact via cytokine signalling \cite{jensen,cytokines}, which is non-specific, and intra-clonal response can be highly cooperative\footnote{In particular, in biochemistry the degree of cooperation is measured by the Hill coefficient $C$ \cite{biochimica,aldo}, such that for high values, e.g. $C \geq 4$ there is a strong sigmoidal shape in the response: The system shows unresponsiveness to small stimuli while it is maximally responding once the stimulus reaches a threshold and stay stable beyond.} so that $h_i$ is better approximated by a step function (a steep hyperbolic tangent). Consequently, we define the status (active/inactive) of helper clones as a ``hard spin'', such that if $h_i=-1$ the $i^{th}$ clone is quiescent, while if $h_i=+1$ it is firing, namely secreting cytokines\footnote{As simplifying assumptions, we neglected details about the subclasses $T_{h_1},T_{h_2}$ \cite{jensen} and we neglected the "hierarchical" strength of various cytokines \cite{hierarchical} (e.g. interferons usually induce  stronger responses w.r.t. interleukins) assuming all the chemical signals as equivalent. These assumptions can be relaxed by assuming a more complex alphabet for chemical messengers.}.

Cytokines are cell-signalling protein molecules able to make B-clones grow or shrink, respectively. Denoting with $\xi_i^{\mu}$ the cytokine exchanged by the $i^{th}$ helper clone and the $\mu^{th}$ B-clone, we confer to these chemical messengers  either positive (expansion) or negative (suppression) signs and we assume them to play as quenched variables, extracted from an a-priori probability distribution.
It should be remarked that the interaction between B and helper cells is rather complex as it requires that B-cells first recognize and engulf a matching antigen and then display antigen fragments bound to their major histocompatibility complex \cite{abbas}. Then, the helpers, attracted by this combination, secrete cytokines directed to B cells themselves.
Here, we generically refer to helper cells as T-cells secreting cytokines, without distinguishing between ``helper'' T-cells (secreting stimulatory cytokines) and "regulatory" or "suppressor"  T lymphocytes (secreting inhibitory cytokines). Indeed, such a discrimination is intrinsic in our model since clones coupled with a negative (suppressive) cytokine can be seen as suppressors, while the clones coupled to positive (eliciting) cytokine can be seen as helpers.

Taking into account all the related stimuli, the evolution of an arbitrary B clone can be described as
\begin{equation}\label{langevin}
\tau\frac{db_{\mu}}{dt}= - \frac1B \sum_{\nu=1}^B I_{\mu\nu}b_\nu+\sqrt{\frac{1}{H}}\sum_{i=1}^H\xi_i^{\mu}h_i+ A_{\mu}+\sqrt{\frac{2\tau}{\beta}}\eta_{\mu}(t).
\end{equation}
In the l.h.s. $\tau$ sets the typical time-scale for the growth of B-clones ($\sim 1$ week), which is described through the time derivative of $b_{\mu}$. In the r.h.s. we have three contributions: the first one accounts for the B-B interactions (see Fig.~\ref{Uovo}, panel $a$); the second one accounts for interactions with helpers through cytokines
and defines the bipartite B-H network (see Fig.~\ref{Uovo}, panel $b$); the third term accounts for a pathogen insult, which is linearly coupled to the corresponding B-clone; the fourth term accounts for a standard white noise with zero mean  and covariance $\langle \eta_{\mu}(t)\eta_{\nu}(t') \rangle = \delta_{\mu,\nu}\delta(t-t')$ whose fluctuation strength is ruled by $\beta$.

For the sake of simplicity, we neglect second-order effects due to the idiotypic network (see e.g. \cite{PRE,noi1,noi2,mario,markus1,markus2}), and we set $I_{\mu \nu}=1, \forall \mu, \nu$. Hence, within such a mean-field approximation we can write $(1/B) \sum_{\nu=1}^B I_{\mu\nu}b_\nu \sim b_{\mu}$, so to get the dynamical evolution of the generic $\mu^{th}$ B-clone as
\begin{equation}\label{PS2}
\tau \frac{d  b_{\mu}}{dt} =  - b_{\mu}  + \sqrt{\frac{1}{H}}\sum_{i=1}^{H} \xi_i^{\mu}h_i + A_{\mu}+\sqrt{\frac{2\tau}{\beta}}\eta_{\mu}(t) = -\frac{d}{d b_{\mu}}\mathcal{H}(h,b;\xi)+\sqrt{\frac{2\tau}{\beta}}\eta_{\mu}(t),
\end{equation}
where, in the last  equality we highlighted the Hamiltonian representation holding as far as the interactions are assumed symmetric. In fact, this dynamics converges to a Boltzmann-like measure on the $\{b,h\}$-phase-space implicitly defined by the following Hamiltonian
\be\label{prima}
\mathcal{H}(h,b;\xi) = \frac12\sum_{\mu=1}^B b_{\mu}^2 - \sqrt{\frac{1}{H}}\sum_{\mu}^B \sum_{i=1}^H \xi_i^{\mu}h_i b_{\mu} - \sum_{\mu=1}^B b_{\mu}A_{\mu}.
\ee
Let us focus on the simpler framework of null external stimuli ($A_{\mu}=0$), and outline the plan: Once an Hamiltonian representation is achieved (Eq. (\ref{prima})), defining $\alpha=B/H$, it is possible to introduce the corresponding partition function $Z_{H,B}(\beta,\alpha)$, as a sum over all the configurational space of the Maxwell-Boltzmann weight $\exp[-\beta \mathcal{H}(h,b;\xi)]$, and from the partition function derive the free energy of the system $F(\beta,\alpha) \propto \ln Z_{H,B}(\beta,\alpha)$. Extremizing the free energy (hence minimizing the energy and maximizing the entropy of the system) then offers the thermodynamics of the model, where spontaneous collective behavior can be observed.

Now, the partition function $Z_{H,B}(\beta,\alpha)$ of this system, defined as
$$Z_{H,B}(\beta,\alpha) \equiv  \sum_{ \{ h \}}^{2^H}\int \prod_{\mu=1}^B d b_{\mu} e^{-\beta \mathcal{H}(h,b;\xi)},$$
can be written as
\begin{eqnarray}\label{summary}
Z_{H,B}(\beta,\alpha)&=& \sum_{ \{ h \}}^{2^H}\int \prod_{\mu=1}^B d b_{\mu} e^{-\sum_{\mu=1}^B \beta \frac{b_{\mu}^2}{2}}
e^{\frac{\beta}{\sqrt{H}} \sum_{i \mu}^{H B}\xi_i^{\mu}h_i b_{\mu}} \\
\nonumber
&=& \sum_{ \{ h \}}^{2^H} \exp\left( \frac{\beta}{2H}\sum_{ij}^H \sum_{\mu=1}^B \xi_i^{\mu}\xi_j^{\mu} h_i h_j \right)=
 \sum_{ \{ h \}}^{2^H} \exp [-\beta \tilde{\mathcal{H}}(h;\xi)],
\end{eqnarray}
where the interactions among B-clones, which constitute the first term in Eq. (\ref{prima}), allow convergence in the integrals of Eq. (\ref{summary}) acting as Gaussian measures.
In the statistical mechanics scaffold, the parameter $\beta$ rules the level of noise in the network; notice that above we mapped $\beta^2$ into $\beta$.

We stress that, interestingly, the complex interactions between helpers and B-cells are absorbed, via marginalization (namely Gaussian integration), within a two-body Hamiltonian $\tilde{\mathcal{H}}(h;\xi)$, namely the evolution of B clones is recast into the thermodynamics of a system of T-clones making up a monopartite, weighted network (see Fig.~\ref{Uovo}, panel $c$), where links are built according to the Hebbian structure
\begin{equation} \label{eq:hebb1}
J_{ij} = \sum_{\mu=1}^B \xi_i^{\mu} \xi_j^{\mu}.
\end{equation}
Indeed, $\tilde{\mathcal{H}}$ turns out to be equivalent to the Hamiltonian of the Hopfield model, whose striking property is that, under proper conditions, it is able to work as an associative memory \cite{amit}: Interpreting the string $\xi^{\mu}$ as a pattern of information and introducing the set of ``pattern overlaps''
\be\label{eq:overalp}
m_{\mu} \equiv \frac{1}{H} \sum_{\mu=1}^B \xi_i^{\mu} h_i,
\ee
one finds that when $\alpha \leq \alpha_c$ and $\beta>\beta_c$ \cite{peter}, the system typically relaxes to a state where one pattern, say the $\mu^{th}$ one, is perfectly retrieved, which means that for any $i$, $h_i = \xi_i^{\mu}$ (under gauge invariance). In this state $m_{\mu}=1$ and $m_{\nu}=0$ for any $\nu \neq \mu$. In fact, overlaps, also called Mattis magnetizations, are order parameters of the model meaning that they are able to quantify the phases of the system as they are zero when the system displays no collective capabilities in retrieval and differ from zero otherwise.

Here, the memorized ``patterns of information" correspond to particular strategies, encoded by cytokine secretions, directed to B-cells.
Therefore, the overlap $m_{\mu}$ related to the $\mu^{th}$ B-clone is larger whenever the signaling from helpers is concerted (T-cells associated to excitatory/inhibitory signals over the B-clone considered are/are-not firing) and this corresponds to an activation of the $\mu^{th}$ B-clone itself. This situation can be looked at as the retrieval of the strategy aimed to expand the $\mu^{th}$ B-clone itself, following, for instance the insult from a matching pathogen.

As mentioned above, the system described by $\tilde{\mathcal{H}}(h;\xi)$ can show cooperative cognitive features as long as $\alpha \leq \alpha_c$, where $\alpha_c \leq 1$ is a critical value implicitly offering the first global constraint for a correct performance of the immune system: Helper T cells must be more than B cells and this fact is indeed confirmed experimentally \cite{abbas}. Interestingly, from this viewpoint the breakdown of immune-surveillance by H.I.V. infection or the (temporary) breakdown due to E.B.V. infection can be associated to an anomalous large value of $\alpha$: in the former case this stems from a drop in the number of helper cells and in the latter from a growth in the number of B cells.

These concepts can be better understood by rewriting he Hamiltonian of the model in terms of pattern overlaps $m_{\mu}$, that is
\be\label{toy2}
\beta\tilde{\mathcal{H}}(h;\xi) = -  \frac{\beta}{2H}\sum_{i, j=1}^H \sum_{\mu=1}^B \xi_i^{\mu} \xi_j^{\mu} h_i h_j = - \frac{\beta H}{2}\sum_{\mu=1}^{B}m_{\mu}^2,
\ee
in such a way that the minimum (free)-energy principle implies $m_{\mu}=1$ for a particular $\mu$ (this is called the ''pure state ansatz" \cite{amit}).
If, for example, the pattern of cytokine activation concerning the B clone $\mu = 1$ has been retrieved, then $m_1=1$, which means that all helper states $h_i$ are parallel to the corresponding cytokines $\xi_i^1$ linking them to the first B-clone. As a consequence, all the inhibitor signals are absent (because each $\xi_k^1=-1$ is coupled to $h_k=-1$ such that their product is positive) while all the eliciting signals are present (because each $\xi_l^1=+1$ is coupled  to $h_k=+1$). Thus, the assembly of helpers spontaneously orchestrates the response against the antigen coupled to the B-clone $\mu=1$, conferring to the  latter the maximal strength for the clonal expansion; from this point classical Burnet theory follows.

The model described so far has been implicitly embedded in a mean-field framework \cite{JTB}, where each helper clone is supposed to interact  with the whole B-repertoire and the assembly for deciding about one single B-clone includes the whole helper ensemble. These are both unrealistic features given the huge sizes of such populations and the fact that interactions are essentially local and of diffusive nature. In what follows we remove the hypothesis of a fully-connected bi-layered spin-glass network, and we allow only a (small) fraction of the whole ensemble of helpers to coordinate the response of a given B-clone.

\section{Getting closer to biology: Dilution in the B-H interactions}

When dilution is absent, the minimization of the free-energy derived from the Hamiltonian (\ref{toy1}) implies\footnote{this holds rigorously for patterns $\vec{\xi}^{ \mu}\cdot\vec{\xi}^{ \nu}=H\cdot \delta(\mu-\nu)$, where orthogonality results  from uncorrelated distributions in the thermodynamic limit $H \to \infty$)} the expansion of the $\mu^{th}$  B-clone (i.e. $m_{\mu} > 0$), but it provides the other clones with no net information (i.e. $m_{\nu \neq \mu} = 0$).
Conversely, real immune systems are able to address a wide variety of antigens simultaneously managing several clones at the same time and, in this sense, we refer to parallel processing capabilities of the network. This property can be restated as the ability to have equilibrium states with several order parameters $m_{\mu}$, $\mu=1,...,K$, different from zero (or above the noise level at finite volume), without being spurious states \cite{amit}. As we are going to show, this property can be captured by systems where couplings are diluted.

More precisely, we introduce dilution in couplings, by writing:
\be\xi_{i \mu}=\varepsilon_{i \mu}\cdot c_{i \mu},\ee
where $\varepsilon_{i \mu}$ assumes values $\pm 1$, representing the excitatory or inhibitory quality of the link (cytokine), and $c_{i \mu}$ assumes values $1$ or $0$ representing existence or absence of the link, respectively.
Their probability distribution are:
\begin{eqnarray}\label{otto}
P(c_{i \mu})&=& d \, \delta_{(c_{i \mu})}+(1-d)\delta_{(c_{i \mu}+1)},\\ \label{nove}
P(\varepsilon_{i \mu}) &=& \frac12 \delta_{(\varepsilon_{i \mu}-1)}+ \frac12 \delta_{(\varepsilon_{i \mu}+1)},
\end{eqnarray}
where $d$ can range continuously in $[0,1]$, allowing some intensive tuning\footnote{The assumption of symmetry for cytokine distribution (see Eq. (\ref{nove})) can be easily relaxed leading to a network with low level of activation consistently with real systems \cite{AGSa}.}.
Hence, we get the following distribution for $\xi_{i \mu}$:
\begin{equation}\label{PB}
P(\xi^{\mu}_{i})=P(c_{i \mu}\varepsilon_{i \mu})=\frac{1-d}{2} \delta_{(\xi^{\mu}_{i}-1)}+ \frac{1-d}{2}\delta_{(\xi^{\mu}_{i}+1)}+  d  \delta_{\xi^{\mu}_{i}},
\end{equation}
such that for $d  \to 1$ no network exists, while for $d \to 0$ the Hopfield model is recovered.

As it is immediate to check, each missing link between the $i^{th}$ T-clone and the $\mu^{th}$ B-clone in the bipartite B-H network appears as a $0$ (i.e. $\xi_i^{\mu}=0)$ in the $i^{th}$ entry of the bit-string $\xi^{\mu}$ in the equivalent associative network, and this ultimately
affects the interaction matrix $\mathbf{J}$ among the helpers.
The following subsections are devoted to the investigation of the properties of the matrix $\mathbf{J}$ and of the weighted graph it generates.

\subsection{Notes about the coupling distribution}

Let us consider a set of $H$ nodes labeled as $i=1,...,H$ and let us associate to each node a string of length $B$ and built from the alphabet $\{-1, 0, 1 \}$, meaning that the generic element $\xi_i^{\mu}$, with $i \in [1,H]$ and $\mu \in [1,B]$, can equal either $\pm 1$ or $0$.
For the H-H network described by the Hamiltonian under investigation, the interaction strength between two arbitrary nodes $i$ and $j$ is given by Eq.~\ref{eq:hebb1}, which is reported here
\begin{equation}\label{eq:heb}
J_{ij} =  \sum_{\mu=1}^B \xi_i^{\mu} \xi_j^{\mu}.
\end{equation}
Of course $J_{ij} \in [-B,B]$. Equation (\ref{eq:heb}) generates a network of mutually and symmetrically interacting nodes, where a link between nodes $i$ and $j$ is drawn whenever they do interact directly ($J_{ij} \neq 0$), either imitatively ($J_{ij}>0$) or anti-imitatively ($J_{ij}<0$).

\begin{figure}[tb] \begin{center}
\includegraphics[width=.6\textwidth]{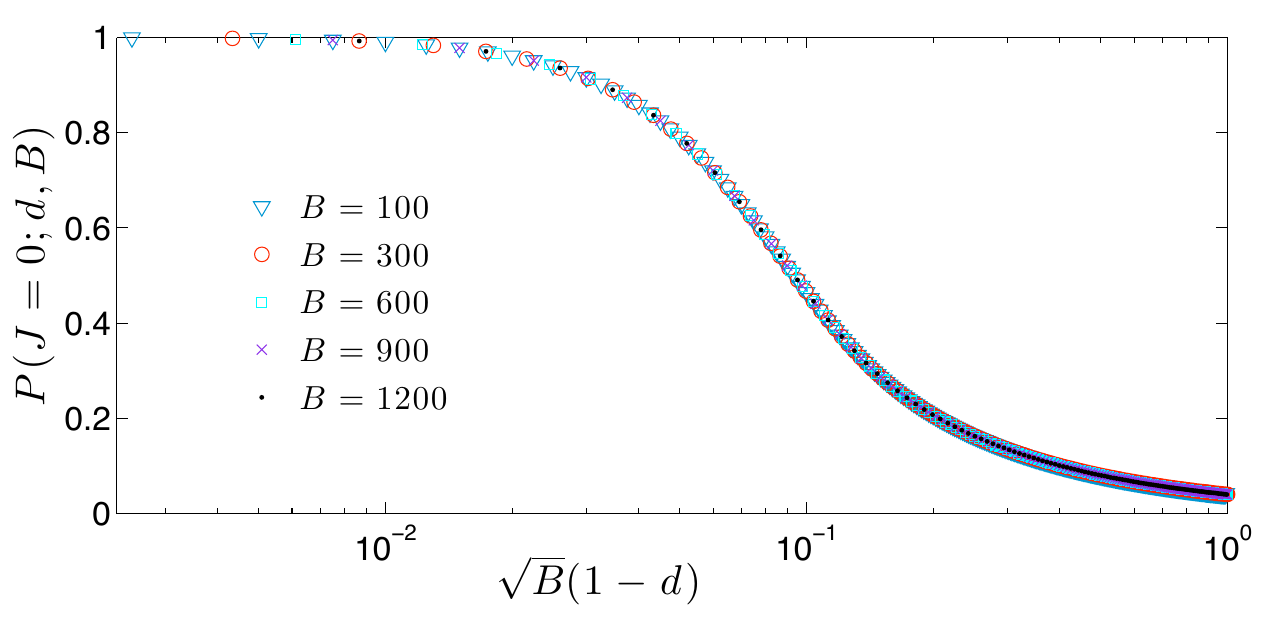}
\caption{\label{fig:PJ0} (Color on line) The probability $P(J=0; d, B)$ is plotted as a function of the dilution $d$ and for different values of $B$, as shown by the legend. Notice the semilogarithmic scale and that dilution is rescaled by $\sqrt{B}$ so to highlight the common scaling of the distributions.}
\end{center}
\end{figure}

First, one can calculate the probability that two nodes (since they are arbitrary we will drop the indices) in the H-H network are linked together, namely
\begin{equation}
P_{\mathrm{link}}(d,B)=P(J \neq 0;d,B) = 1 - P(J=0;d,B) = 1 - \sum_{k=0}^B P_{\mathrm{sum-0}}(k;d,B),
\end{equation}
where $P_{\mathrm{sum-0}}(k; d, B)$ is the probability that two strings display (an even number) $k$ of non-null matchings summing up to zero; otherwise stated, there exist exactly $k$ values of $\mu$ such that $\xi_i^{\mu} \xi_j^{\mu} \neq 0$ and they are half positive and half negative.
In particular, $P_{\mathrm{sum-0}}(0; d, B) = [d(2-d)]^B$, because this is the probability that, for any $\mu \in [1, B]$, at least one entry (either $\xi_i^{\mu}$ or $\xi_j^{\mu}$ or both) is equal to zero. More generally,
\begin{equation}
P_{\mathrm{sum-0}}(k;d,B) = \left(\frac{1-d}{2}\right)^{2k} [d(2-d)]^{B-k}  \binom{B}{k}  \left[2^k \binom{k}{k/2}\right],
\end{equation}
where the first and the second factors in the r.h.s. require that $k$ entries are non-zero and the remaining $B-k$ entries are zero; the third factor accounts for permutation between zero and non-zero entries, while the last term is the number of configurations leading to a null sum for non-null entries.
Therefore, we have
\begin{equation}
P(J=0;d,B) =  [d(2-d)]^{B} \sum_{k=0}^B  \left[ \frac{(1-d)^2}{2d(2-d)}\right]^{k}  \binom{B}{k}  \binom{k}{k/2},
\end{equation}
whose plot is shown in Fig.~\ref{fig:PJ0}. As for its asymptotic behavior, we can expand for $d$ close to $1$ and close to $0$ (for simplicity we assume $B$ finite and even) getting, respectively,
\begin{eqnarray}\label{eq:Dgrande}
P(J=0;d,B) &=& 1 - B (1-d)^2 + \frac{3}{4} B(B-1)(1-d)^4+\mathcal{O}(1-d)^6\\
\label{eq:Dpiccolo}
P(J=0;d,B) &=& \frac{(-1)^{B/2} \sqrt{\pi}}{\Gamma(1/2-B)\Gamma(1+B/2)} \left( 1 -2B \, d \right)  + \mathcal{O}(d^{2}) \\
\nonumber
&\approx& \frac{1-2B\,d}{4^{B/2}} \binom{B}{B/2} + \mathcal{O}(d^{2}).
\end{eqnarray}
The average number of nearest neighbors per node $\langle z \rangle_{d,B,H}$ follows immediately as $\langle z \rangle_{d,B,H} = H  P_{\mathrm{link}}(d,B)$.

More generally, we can derive the coupling distribution $P(J;d,B)$, once having defined $P_{+1}(k)$, $P_{-1}(k)$ and $P_{0}(k)$, as the probability that, given two strings, they display $k$ matches each equal to $+1$, $-1$ and $0$, respectively, namely
\begin{equation}
P_{+1}(k;d)= P_{-1}(k;d)=\left[ \frac{(1-d)^2}{2} \right]^k, \ \ P_0(k;d) = \left[d(2-d) \right]^k.
\end{equation}
Hence, we can write
\begin{eqnarray}\label{eq:gaussian}
P(J;d,B)&=& \sum_{l=0}^{(B-J)/2} P_{+1}(l+J;d)P_{-1}(l;d)P_0(B-2l-J;d)\\
&\times& \frac{B!}{l!(l+J)!(B-2l-J)!} \sim \mathcal{N}(0,\sigma_J(d,B)). \nonumber
\end{eqnarray}
The last asymptotic holds for large $B$, hence it is sufficient that $B$ scales like $H^{\gamma}$, with $\gamma \leq 1$ in the limit $H \to \infty$. The null mean value $\langle J \rangle_{d,B} =0$ is due to the symmetry characterizing $P(\xi_i^{\mu})$ (see Eq. ($9$)), while the standard deviation is $\sigma_J(d,B) = \sqrt{\langle J^2 \rangle_{d,B}} = \sqrt{B} (1-d)$.


It is worth underlining that $P(J;d,B)$ does not depend on the size $H$. Indeed, patterns are drawn independently and randomly so that the coupling $J_{ij}$ may be regarded as the distance covered by a random walk of length $B$ and endowed with a waiting probability $d(2-d)$. Hence, the end-to-end distance is distributed normally around zero and with variance (mean squared distance) given by the effective number of steps performed, according to the diffusion law, namely $\sim [1-d(2-d)]B=(1-d)^2 B$, in agreement with results above.

\subsection{Pattern dilution versus Topological dilution}
When dilution on pattern entries is introduced, a topological dilution in [H-H] can be induced, and, as we will see, the resulting structure is far different from the one which would be realized by a random bond deletion. 
Even from a thermodynamic point of view, the behavior of the diluted system is deeply different from the case of a Hopfield model where edges are randomly deleted \cite{Agliari-PRL2012}. 

Let us first focus on the topological properties of the emerging monopartite graph.
First, we recall that, according to a mean-field approach, the network is expected to display a giant component  when the average link probability is larger than $1/H$.
In the thermodynamic limit and assuming a large enough size $B$ (stemming from either low, i.e. $B \sim \log H$, or high, i.e. $B \sim H$, storage regimes) to ensure the result in Eq. (\ref{eq:gaussian}) to hold, for any finite value of $1 - d$ the emergent graph turns out to be always overpercolated. In fact, $P_{\textrm{link}}(d,B) = 1-P(J=0; d,B) \sim 1 - 1/ \sqrt{2 \pi \sigma_J^2}$, so that it suffices that $\sigma_J > H/[\sqrt{2 \pi} (H-1)] \to 1 / \sqrt{2 \pi}$ and this leads to $d<1-(2 \pi B)^{-1/2} \to 1$.

Similarly, when $B$ is finite we can check the possible disconnection of the network by studying $P(J=0;d,B)$ from Eq. (\ref{eq:Dgrande}) and we get that $P_{\textrm{link}}(d,B)<1/H$ for $d > 1 - 1/{\sqrt{BH}}$. Thus, in the thermodynamic limit, for any finite value of $1-d$, the graph is still overpercolated. Replacing $1/H$ with $(\log H)/H$, one also finds that the graph is even always connected.

\begin{figure}[tb] \begin{center}
\includegraphics[width=4.2cm, angle=0]{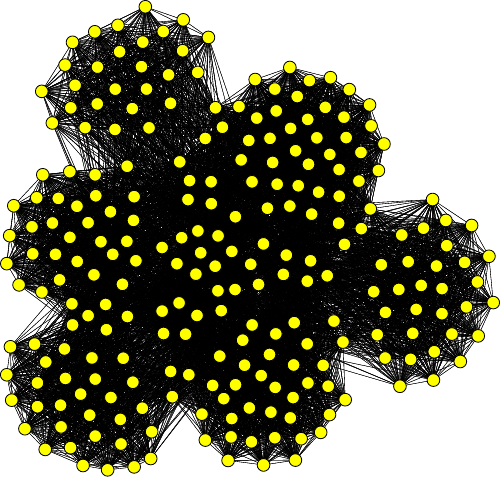}
\includegraphics[width=4.0cm, angle=0]{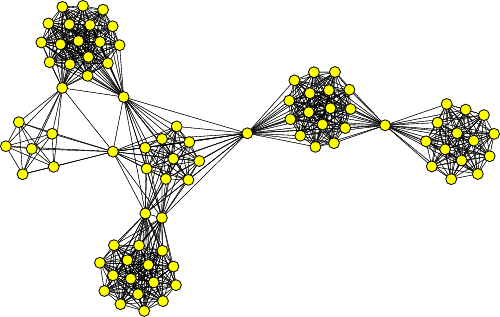}
\includegraphics[width=4.2cm, angle=0]{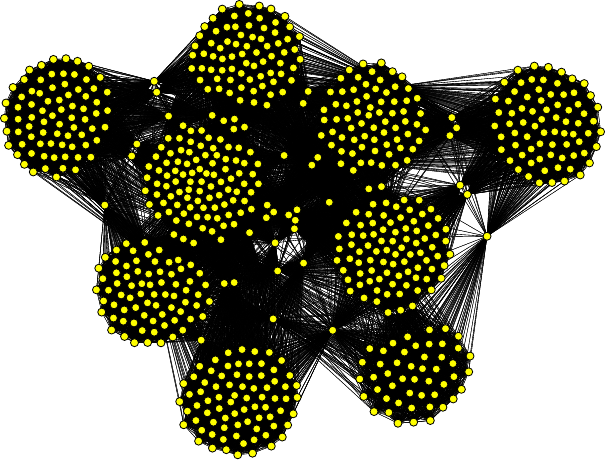}
\caption{\label{fig:graphs} (Color on line) Examples of structures for the [H-H] system obtained for relatively high values of dilution and different sizes $H$ and $B = \log (H)$, namely $H=500$, $B=6$ and $d=0.85$ (leftmost panel); $H=500$, $B=6$ and $d=0.98$ (middle panel); $H=5000$, $B=8$ and $d=0.98$ (righmost panel); isolated nodes are omitted from these plots. The arrangements of nodes have been realized so to highlight the modularity of the structures which emerges especially at large dilutions. A node $i$ working as a bridge betwen modules correspond to a string $\xi_i$ displaying more than one non-null entries.}
\end{center}
\end{figure}

In Fig.~\ref{fig:graphs} we report some examples of structures for several choices of parameters; these evidence that when $d$ is (relatively) small, the graph is (almost) fully-connected, while, as $d$ gets close to $1$, the graph starts to exhibit high modularity with a number of cliques equal to $B$. More precisely, nodes corresponding to helper clones and that in the bipartite graph [B-H] are connected to the same node, say $\mu$, will form in the monopartite graph [H-H] a clique; a node that in the bipartite graph has $k$ neighbors, in the monopartite graph will serve as a bridge between $k$ cliques (see also \cite{PRE,cioli}).

If we take $d$ properly approaching to $1$ as $H$ is increased, different scenarios may emerge  \cite{EPL,BA}.

Another kind of dilution can be realized by directly cutting edges in the resulting associative network, as for instance early investigated in the neural scenario by Sompolinsky on the Erd\"{o}s-Renyi graph \cite{sompolinsky,amit} or more recently by Coolen and coworkers on small worlds and scale-free structures \cite{ton1,ton2}.
Such different ways of performing dilution - either on links of the associative network (see \cite{sompolinsky,amit,ton1,ton2}) or on pattern entries (see Eq. (\ref{PB})) - yield deeply different thermodynamic behaviors, as for instance evidenced in \cite{Agliari-PRL2012}, by looking at the distribution of fields insisting on each spin, namely for the generic $i^{th}$ spin $\varphi_i = \sum_{i \neq j=1}^N J_{ij} \sigma_j$.

\section{Parallel processing performances}

\subsection{Statistical mechanics of the low-storage case}

As a minimal bibliography in the statistical mechanics approach, we report that a different study sharing some similarities with ours, investigates an associative network with pattern inhibition (due to chemical modulation) in the neuroscience scenario \cite{acc1,acc2}, while a macroscopic behavior close to parallel processing was reported in \cite{acc3}, where more than one overlap were able to retain strictly positive values owing to strong pattern correlations (a completely different motivation with respect to ours).

Now, we solve the model in the low storage regime $B \sim \log H$, such that the limit $\alpha=\lim_{H \to \infty} B/H =0$ holds\footnote{Results outlined within this scaling can be extended with little effort to the whole region $B \sim H^{\gamma}$, with $\gamma<1$, such that the constraint $\alpha=0$ is preserved, as realized in the Willshaw model \cite{will} concerning neural sparse coding.}.
Like in the Amit-Gutfreund-Sompolinsky (AGS) neural network \cite{amit}, the comprehension of the non-saturated case ($\alpha=0$) is the first fundamental step to face before moving to the saturated case ($\alpha>0$). This can be accomplished in several ways and here we adopt the approach described in \cite{peter}.

As standard in disordered statistical mechanics, we introduce three types of average: the Boltzmann average $\omega(m_{\mu}) = \sum_{h} m_{\mu} \exp(-\beta \tilde{\mathcal{H}}(h;\xi))/Z_{H,B}(\beta,d)$, the average $\mathbb{E}$ performed over the quenched disordered couplings $\xi$, and the global expectation $\mathbb{E}\omega(m_\mu)$ defined by the brackets $\langle m_\mu \rangle_{\xi}$.

The equilibrium equations for the order parameter can be obtained from the quenched free energy $\langle F(\beta,d)\rangle_{\xi}$ defined as
\be
\langle F(\beta,d) \rangle_{\xi}=-\lim_{H \to \infty}\frac{1}{\beta}\mathbb{E}\log Z_{H,B}(\beta,d)=-\lim_{H \to \infty}\frac{1}{\beta}\mathbb{E}\log\sum_{\{h \}}^{2^H}e^{-\beta \tilde{\mathcal{H}}(h;\xi)}.
\ee
Introducing the notation $\mathbf{m}=(m_1,...,m_B)$ and $\mathbf{\xi_i}=(\xi_i^1,...,\xi_i^B)$, the above equation can be expressed in terms of the density of states $\mathcal{D}(\mathbf{m})$
\be
\mathcal{D}(\mathbf{m})\equiv \sum_{\{h \}}^{2^H}\delta(\mathbf{m}-\mathbf{m}(h)),
\ee
as
$$
Z_{H,B}(\beta,d) =  \int d\mathbf{m} Z(\mathbf{m}), \ \ \ Z(\mathbf{m}) = e^{H\beta \mathbf{m}^2/2}\mathcal{D}(\mathbf{m}).
$$
Notice that the delta function here is a product of independent delta functions, one for each B-clone, namely:
$$
\delta\left(\mathbf{m}-\mathbf{m}\left(h \right)\right) = \prod_{\mu=1}^B \delta \left(m_{\mu}- m_{\mu}\left(h \right) \right).
$$
We need now to introduce $B$ integration variables $\mathbf{x}=(x_1,...,x_B)$ to switch the delta functions to their integral representation as
$$
\mathcal{D}(\mathbf{m}) = \left( \frac{H}{2\pi} \right)^B\int d \mathbf{x} e^{i H \mathbf{x} \cdot \mathbf{m}}\sum_{\{ h\} }^{2^H}e^{-i\sum_i^H \sum_{\mu}^B h_i \xi_i^{\mu}x_{\mu}} =
\left( \frac{H}{2 \pi} \right)^B\int d \mathbf{x} e^{H \left( i \mathbf{x} \cdot \mathbf{m}+ \langle
\ln 2 \cos(\mathbf{x}\cdot \mathbf{\xi}) \rangle_{\xi} \right)},
$$
where we assumed the property $\lim_{H \to \infty}\sum_i^H f(\mathbf{\xi_i})/H=\langle f(\mathbf{\xi})\rangle_{\xi} $.
\newline
Physically speaking, the log-density of the states quantifies the constrained entropy $S(\mathbf{m})$ and can be evaluated through saddle point integration because of the factor $H$ in the exponent of its integral representation above. Strictly speaking, we calculate only the leading term of the density of states, which is the one retaining statistical meaning in the thermodynamic limit and it is given by the maximum over $\mathbf{x}$ of $S(\mathbf{x},\mathbf{m})$, the latter being
$$
S(\mathbf{x},\mathbf{m}) = i \mathbf{x} \cdot \mathbf{m} + \langle
\ln 2 \cos(\mathbf{x}\cdot \mathbf{\xi}) \rangle_{\xi}.
$$
It is then clear that the intensive quenched free energy  can be rewritten as
\be
\lim_{H \to \infty} \langle F(\beta,d)/H \rangle_{\xi} = -\frac{1}{\beta}\log 2-  \lim_{H \to \infty} \frac{1}{H\beta}\int d\mathbf{m}\mathcal{D}(\mathbf{m})e^{\frac{1}{2}\beta H \mathbf{m}^2}.
\ee
The main contribution to free-energy can be made explicit as a finite-dimensional integral;
as outlined before for the constrained entropy, through the extensively linearity property of thermodynamic observables, for large values of $H$ the integral will be dominated by the saddle-point that maximizes the exponent as
\begin{eqnarray}
\lim_{H \to \infty} \langle F(\beta, d)/H\rangle_{\xi} &=& -\lim_{H \to \infty} \frac{1}{H\beta}\int d\mathbf{m} \, d\mathbf{x}e^{-H\beta f(\mathbf{x},\mathbf{m})}= \textrm{extr} [f(\mathbf{x},\mathbf{m})], \\
f(\mathbf{x},\mathbf{m}) &=& -\frac{1}{2}\mathbf{m}^2-i\mathbf{x}\cdot \mathbf{m}-\frac{1}{\beta}\langle \log 2\cos[\beta \mathbf{\xi}\cdot \mathbf{x}]\rangle_{\mathbf{\xi}}.
\end{eqnarray}
To identify the various ergodic components (which are expected to be $B+1$, one being the paramagnetic one) we find the stationary points of $f(\mathbf{m})$ through the system $\partial_{m_{\mu}} f (\mathbf{m})=0$ for all $\mu \in (1,...,B)$, which gives the vectorial self-consistence equations
\be
\mathbf{x}=i\beta\mathbf{m}, \,\,\,\,\,\,\,\,\,\,\,\,\,\, i\mathbf{m}=\langle\mathbf{\xi}\tan[\mathbf{\xi}\cdot \mathbf{x}]\rangle_{\mathbf{\xi}}.
\ee
Being the saddle point values of $\mathbf{x}$ purely imaginary, and using $\tanh(x)=-i\tan(ix)$ we get
\be\label{alpha0}
\mathbf{m}=\langle\mathbf{\xi}\tanh[\beta \mathbf{\xi}\cdot \mathbf{m}]\rangle_{\mathbf{\xi}}.
\ee
Then, the above equation has to be averaged over the pattern distribution $P(\xi_i^{\mu})$ and finally solved numerically, as explained in the examples of Secs.~\ref{sec:B2} and \ref{sec:B3}.

Before proceeding it is worth noticing that the Hamiltonian $\tilde{\mathcal{H}}(h;\xi)$ of Eq. (\ref{toy2}) is quadratic in the pattern overlaps $m_{\mu}$ and the $B$ stored patterns contain (on average) a fraction $d$ of null entries. As a consequence, the pure state ansatz $(m_1=1,m_2=...=m_B=0)$ \cite{amit} can no longer work.
In fact, now, the retrieval of a pattern (say $\xi^1$, the one coupled to $m_1$) does not employ all the available spins (and coherently $m_1 < 1$, for $d \neq 0$) and those corresponding to null entries  can be used to recall further patterns up to the exhaustion of all spins.

In particular, at zero noise level and relatively low degree of dilution ($d<d_{c}$), one pattern, say $\mu=1$, is perfectly retrieved, while a fraction $d$ of spins is still available and its overlap with any remaining pattern is, on average, $1-d$; hence, the second best-retrieved pattern, say $\mu=2$, displays a (thermodynamical and quenched) average of the Mattis magnetization equal to $d(1-d)$. In other words, once $m_1$ has been retrieved, it is energetically convenient for the system to coordinate its free helpers to align with another pattern instead of letting them align randomly. Proceeding analogously, one finds
\be \label{eq:ansatz}
m_k=d^{k-1} (1-d).
\ee
Therefore, the overall number $K$ of retrieved patterns corresponds to $ \sum_{k=0}^{K-1} (1-d) d^k =1$, with the cut-off at finite $N$ as, due to discreteness, $(1-d)d^{K-1} \geq N^{-1}$ must hold. For any fixed and finite $d$, this implies $K \lesssim \log N$, which can be thought of as a ``parallel low-storage'' regime of neural networks.
%
Such a hierarchical fashion for alignment, providing an overall energy $- H/2 \sum_k [(1-d)d^k]^2 = - H (1-d^{2+2B})(1-d) / [2(1+d)]$, is more optimal than a uniform alignment of spins amongst the available patterns which would yield $m_k=d/B$ for any $k$ and an overall energy $-H/2 \sum_k (d/B)^2 = -(d^2H)/(2B)$.

On the other hand, at larger degrees of dilution ($d > d_{c}$) and $B>2$, the state (\ref{eq:ansatz}) is no longer stable since no magnetization is large enough to yield a field $\xi_i^{\mu} m_{\mu}$ able to align all the related ($\xi_i^{\mu}  \neq 0$) spins; as a result, the system falls into a spurious state where all patterns are partially retrieved, but none exactly.

The state corresponding to Eq.~\ref{eq:ansatz} can be formally written as
\be
h_i=\xi_i^1+\sum_{\nu=2}^B \xi_i^{\nu}\prod_{\mu=1}^{\nu-1}\delta(\xi_i^{\mu}),
\ee
and it ceases to be stable when $m_1 \leq \sum_{k>1} m_k$. 
For $B>2$ this inequality has a solution, which corresponds to a critical dilution $d_{c}$. It is easy to see that $d_c$ approaches (exponentially from above) $1/2$ in the limit of large $B$ \cite{Agliari-PRL2012}.

The picture described above is corroborated by the numerical solution of Eq.~\ref{alpha0} and by the numerical simulations presented in the following sections.

\subsection{The case $B=2$} \label{sec:B2} 
Despite the structure of the self-consistencies for an arbitrary value of $B$ (\ref{alpha0}) are extremely simple both conceptually and analytically, they become, already for $B >3$, of prohibitive length and handleable only via calculators.
Here, we first focus on the simplest case $B=2$, where the parallel ansatz (\ref{eq:ansatz}) is always stable (see also Appendix) and no spurious state emerges. The analysis of this special case is useful in order to introduce the statistical-mechanics arguments and as a starting point to see how parallel processing does work.
The self-consistencies encoded into Eq. (\ref{alpha0}) for the simplest case $B=2$ read off as
\be\label{kk}
\langle m_1 \rangle_{\xi} = d(1-d)\tanh(\beta \langle m_1 \rangle_{\xi})+\frac{(1-d)^2}{2}\{\tanh[\beta(\langle m_1 \rangle_{\xi} + \langle m_2 \rangle_{\xi})]+\tanh[\beta( \langle m_1 \rangle_{\xi} - \langle m_2\rangle_{\xi})] \},
\ee
\be\label{kkk}
\langle m_2 \rangle_{\xi} =d(1-d)\tanh(\beta \langle m_2 \rangle_{\xi})+\frac{(1-d)^2}{2}\{\tanh[\beta(\langle m_1 \rangle_{\xi} + \langle m_2\rangle_{\xi})]-\tanh[\beta(\langle m_1 \rangle_{\xi} - \langle m_2\rangle_{\xi})] \}.
\ee
The solution of these equations for different values of $\beta$ is reported in Fig.~$6$.
\begin{figure}[tb] \begin{center}
\includegraphics[width=.6\textwidth]{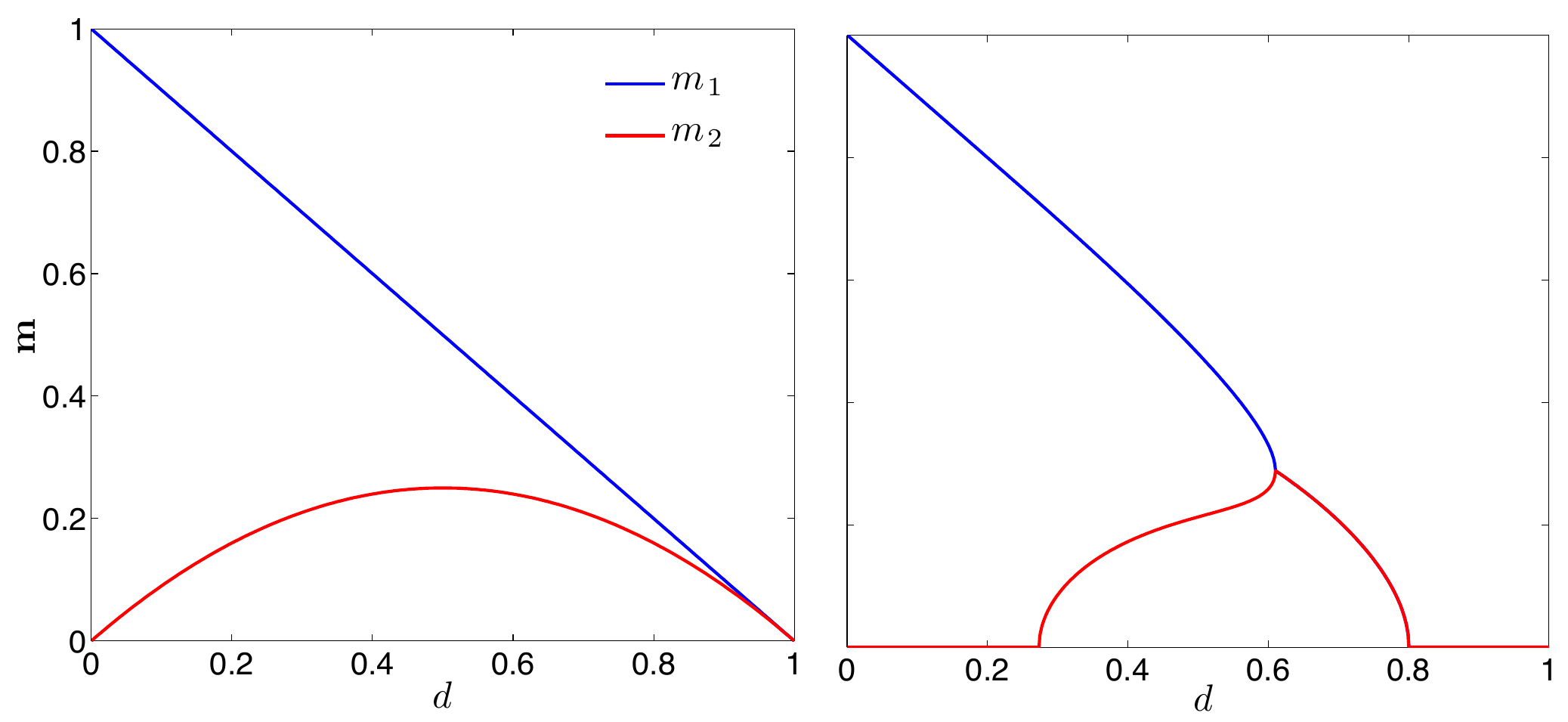}
\caption{\label{fig:M1M2} (Color on line) Behavior of the two Mattis magnetizations $m_1$ and $m_2$ versus $d$ at two (small) noise levels, namely $\beta^{-1}=10^{-4}$ (left panel) and $\beta^{-1}=0.20$ (right panel). We recall that $m_1$ and $m_2$ represent, from a biological perspective, the extent of the signal received by the clones $\mu=1$ and $\mu=2$, respectively. Thus, when both $m_1$ and $m_2$ are relatively large, both clones are prompt to react simultaneously.}
\end{center}
\end{figure}
In the low (fast) noise limit ($\beta \to \infty$, right panel), when no dilution is present ($d=0$) the second magnetization $m_2$ disappears and the first magnetization $m_1$ approaches the value $1$ as expected because the Hopfield model is recovered. As dilution is increased, $m_1$ decreases linearly, while $m_2$ displays a parabolic profile with peak at $d=1/2$. In the presence of (fast) noise (right panel), $m_2$ starts growing for higher values of dilution because (see next subsection and the signal-to-noise analysis of Appendix for further details) the signal
insisting on the latter, which is proportional to $d(1-d)$, must be higher than the noise level in order to be effective. Also notice that, from intermediate dilution onwards, $m_1$ and $m_2$ collapse and the related curves converge at a ``bifurcation'' point.

We now deepen these results, first from a more intuitive point of view, and later from a more rigorous one.

Let us divide spins into four sets: $\mathcal{S}_1$, which contains spins $i$ corresponding to zero entries in both patterns ($\xi_i^{1} = \xi_i^{2}=0$), therefore behaving paramagnetically;  $\mathcal{S}_2$, which includes spins seeing only one pattern ($|\xi_i^{1}| \neq |\xi_i^{2}|$);
$\mathcal{S}_3$, which contains spins corresponding to two parallel, non-null entries ($\xi_i^{1} = \xi_i^{2} \neq 0$), thus being the most stable; $\mathcal{S}_4$, which includes spins $i$ corresponding to two parallel, non-null entries ($\xi_i^{1} = - \xi_i^{2} \neq 0$), hence intrinsically frustrated.
\newline
The cardinality of these sets are: $|\mathcal{S}_1| = d^2$, $|\mathcal{S}_2| = 2d(1-d)$, $|\mathcal{S}_3| = (1-d)^2/2$, and $|\mathcal{S}_4| = (1-d)^2/2$.
Now, the most prone spin to align with the related patterns are those in $\mathcal{S}_3$ and in $\mathcal{S}_2$, and this requires $(1-d) < \beta^{-1}$ for the field to get effective. As $d$ is further reduced, $m_1$ and $m_2$ grow paired, due to the symmetry of the sets $\mathcal{S}_2$ and $\mathcal{S}_3$. The growth proceeds paired until the magnetizations get the value $m_1 = m_2 = (1-d)^2/2 + d(1-d)$, where the two contributes come from spins aligned with both patterns and with the unique pattern they see, respectively.
From this dilution onwards frustrated spins also start to align so that one magnetization necessarily prevails over the other.
This explanation can be extended to any finite $B$ and, in general, the number of sets turns out to be $P+1+\sum_{k=0}^B \lfloor \frac{P-k}{2} \rfloor$.

Now we want to quantify bifurcation points, and to this task let us call
\be
x=\langle m_1 \rangle_{\xi} - \langle m_2 \rangle_{\xi}.
\ee
We use Eqs.~(\ref{kk}) and (\ref{kkk}) and expand for small values of $x$
\be
\nonumber
\langle m_1 \rangle_{\xi} - \langle m_2 \rangle_{\xi} =x=d(1-d)[\tanh(\beta \langle m_1 \rangle_{\xi})-\tanh(\beta \langle m_2 \rangle_{\xi})]+(1-d)^2\tanh\left[\beta( \langle m_1 \rangle_{\xi} - \langle m_2\rangle_{\xi}) \right],
\ee
where
\begin{eqnarray}
\nonumber
&&d (1-d)\left[\tanh\left(\beta \langle m_1 \rangle_{\xi} \right)-\tanh\left(\beta \langle m_2 \rangle_{\xi} \right)\right] \\
\nonumber
&\sim& d (1-d)\left[\tanh(\beta \langle m_1 \rangle_{\xi})-\tanh(\beta \langle m_2 \rangle_{\xi})+\frac{\beta x}{\cosh^2(\beta \langle m_1 \rangle_{\xi})}\right],
\end{eqnarray}
and
\be
\nonumber
(1-d)^2\tanh(\beta \langle m_1 \rangle_{\xi} - \langle m_2 \rangle_{\xi}) \sim (1-d)^2 \beta x + \mathcal{O}(x^3).
\ee
Thus, the leading term is
\be
x \sim \bigg[\frac{d (1-d)\beta}{\cosh^2(\beta \langle m_1 \rangle_{\xi})}+\beta (1-d)^2 \bigg]x.
\ee
The critical value of $\beta$ corresponding to the bifurcation point is defined as
\be
\beta_c^{bif}=\frac{1}{(1-d)^2 \bigg[1+\frac{(1-d)}{d}\frac{1}{\cosh^2(\beta_c^{bif} \langle m_1 \rangle_{\xi}) }\bigg]}.
\ee
This mechanism can be easily generalized to the case $B>2$.

We now analyze the critical noise level at which the order parameters disappear and the network dynamics becomes ergodic: Expanding expressions (\ref{kkk}) we find
\begin{eqnarray}\nonumber
\langle m_2 \rangle_{\xi} &\sim& d (1-d)[\beta \langle m_2 \rangle_{\xi}] +  \frac{(1-d)^2}{2}\Big[\beta \langle m_1 \rangle_{\xi}  +\beta \langle m_2 \rangle_{\xi} \\
\nonumber
&+& \frac{\beta^3}{3}(\langle m_1 \rangle_{\xi}^3+ \langle m_2\rangle_{\xi}^3+3 \langle m_1\rangle_{\xi}^2 \langle m_2 \rangle_{\xi} +3 \langle m_1 \rangle_{\xi} \langle m_2 \rangle_{\xi}^2)\Big]+ \\
\nonumber
&+& d(1-d)\frac{\beta^3}{3} \langle m_2 \rangle_{\xi}^3 -\frac{(1-d)^2}{2}\Big[\beta \langle m_1 \rangle_{\xi} -\beta \langle m_2 \rangle_{\xi}\\
&+&\frac{\beta^3}{3}(\langle m_1\rangle_{\xi}^3- \langle m_2 \rangle_{\xi}^3-3 \langle m_1 \rangle_{\xi}^2 \langle m_2 \rangle_{\xi} + 3 \langle m_1 \rangle_{\xi} \langle m_2\rangle_{\xi}^2)\Big], \nonumber
\end{eqnarray}
such that we can write
\be
\langle m_2 \rangle_{\xi} \sim (1-d) \beta \langle m_2 \rangle_{\xi} + \mathcal{O}(\langle m_2 \rangle_{\xi}^3).
\ee
Therefore the critical noise level turns out to be
\be
\beta_c=\frac{1}{1-d},
\ee
which collapses to $\beta_c=1$ when $d \to 0$ (correctly recovering the Hopfield scenario).
\newline
Again, this calculation can be easily generalized to the case $B>2$, so to find (possibly on a calculator) all the bifurcation points at each desired storage level.

\subsection{The case $B>2$}\label{sec:B3}
In this subsection we present some results for the general case $B>2$.

\begin{figure}[tb] \begin{center}
\includegraphics[width=.55\textwidth]{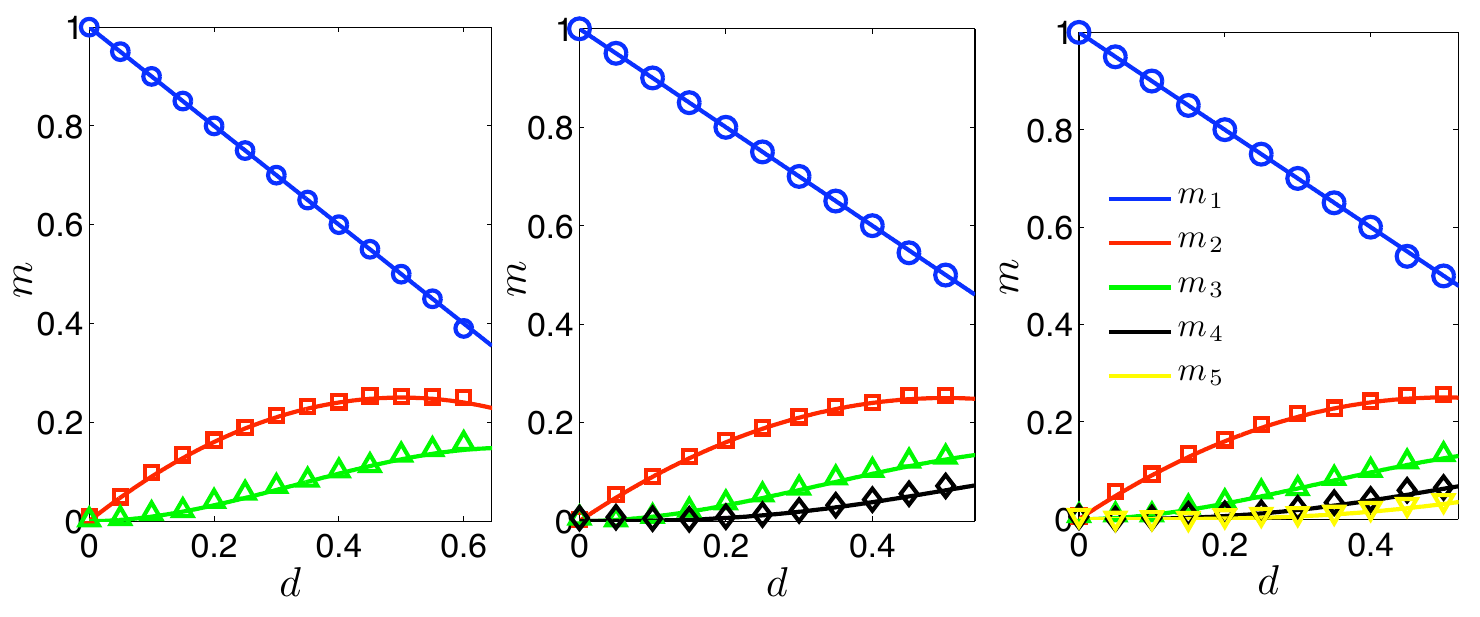}
\caption{\label{fig:345pat} (Color on line) Mattis magnetizations $\mathbf{m}$ versus dilution $d$ for $B=3$ (leftmost panel), $B=4$ (middle panel) and $B=5$ (rightmost panel) patterns at zero noise level ($\beta^{-1}=0$). The degree of dilution ranges in $[0, d_c(B)]$, where $d_c$ depends on the number of patterns considered, namely $d_c \approx 0.61$, $d_c \approx 0.54$, $d_c \approx 0.52$, respectively.
Data from Monte Carlo simulations (symbols) are successfully compared with the analytic results of Eq.~\ref{eq:ansatz} (solid lines).
}
\end{center}
\end{figure}

In general, at zero noise level and for relatively small values of $d$, the parallel ansatz of Eq.~\ref{eq:ansatz} holds, as shown in Fig.~\ref{fig:345pat} where several values of $B$ are considered.

\begin{figure}[tb] \begin{center}
\includegraphics[width=.55\textwidth]{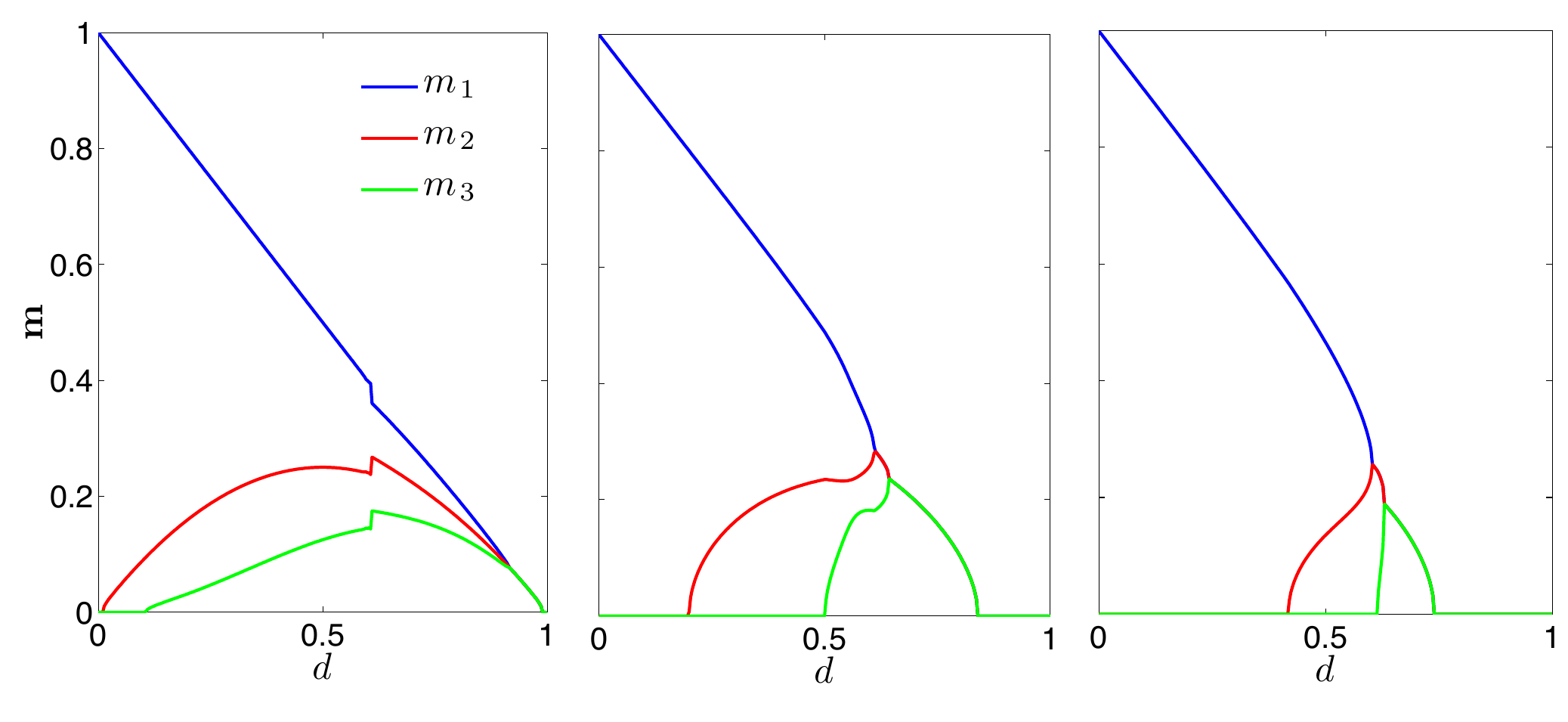}
\caption{\label{fig:3patT} (Color on line) From left to right: Mattis magnetizations $\mathbf{m}$ versus dilution $d$ for $B=3$ at $\beta^{-1}=0.001$, $\beta^{-1}=0.11$ and $\beta^{-1}=0.26$, respectively. The discontinuity occurring at $d =d_c \approx 0.61$ in the leftmost panel corresponds to the failure of the parallel ansatz (\ref{eq:ansatz}): for $d>d_c$ the system relaxes towards a state where none of the patterns is completely retrieved. For large degrees of noise the discontinuity is smoothed out.
}
\end{center}
\end{figure}

When noise is also introduced, we have that for the $k^{th}$ pattern to be retrieved the related field $\xi_i^{k} m_k$ insisting on the $i^{th}$ spin has to be larger than the noise level, that is $[d^{k-1}(1-d)]>\beta^{-1}$, if this condition is not fulfilled the field is confused with the noise and the pattern can not be retrieved.
In particular, $m_2$ is non vanishing only for $d>(1 - \sqrt{1-4/\beta})/2$, $m_3$ is non vanishing only for (approximately) $d> 1 - \beta^{-1} - 2\beta^{-2}$ and so on.
On the other hand, when $d>1-\beta^{-1}$ no pattern is retrieved.
This is confirmed by Fig.~\ref{fig:3patT} (left panel) where the case $B=3$ is considered at different temperatures.

\subsection{The space of configurations}

In this subsection we deepen the structure of parallel states in the configurational space. To this task let us fix a pattern $\xi^1_i$, with $i=1,...,H$, and a dilution $d$, in such a way that $H\,d$ entries of $\xi^1$ are expected to be null and the remaining $H(1-d)$ are expected to be half equal to $+1$ and half equal to $-1$. The number of spin configurations displaying maximum overlap with $\xi^1$ corresponds to the degeneracy induced by null entries, namely $2^{Hd}$; all these configurations lay in an energy minimum because their pattern overlap is maximum (actually the same holds for the symmetrical configurations due to the gauge symmetry of the model).

Let us now generalize this discussion by introducing the number of configurations $n(m,d)$ whose overlap with the given pattern displays $m$ misalignments, in such a way that $n(m,d)$ is given not only by the degeneracy induced by null entries, but also by the degeneracy induced by the choice of $m$ entries out of $H(1-d)$ which have to be mismatched. It is easy to see that $n(m,d)=2^{Hd}\binom{H(1-d)}{m}$. Interestingly, for such configurations the signal felt by a spin $i$ can be written as $\varphi_i = \xi_i^{1}[H((1-d))-2m]$ and the effect of the correction due to the $m$ misalignments might be vanishing in the presence of a sufficiently large level of noise, so that the system is not restricted to the $2^{Hd}$ configurations corresponding to the minimum energy, but it can also explore all the configurations $n(m,d)$. 

Therefore, we can count the number of configurations $\tilde{n}(x,d)$ exhibiting a number of misalignments, with respect to $\xi^1$, up to a given threshold $x$;  in the presence of noise such configurations are all accessible, namely they all lay in the same ``deep'' minimum.
Indeed, we can write $\tilde{n}(x,d) = \sum_{m=0}^x n(m,d)$; of course, for $x=H(1-d)$ we recover $\tilde{n}(x,d)=2^H$.
Moreover, when $x=H(1-d)/2$, we can exploit the identity $\sum_{k=0}^i \binom{2i}{k} = 1/2 [4^i + \binom{2i}{i}]$ \cite{gradshteyn}, and assuming without loss of generality $H(1-d)$ to be even we get
\begin{equation}\label{eq:threshold}
\tilde{n}(H(1-d)/2,d) = \sum_{m=0}^x n(m,d) = \frac{2^{Hd}}{2} \left[ 2^{H(1-d)} + \binom{H(1-d)}{H(1-d)/2} \right] \approx \frac{2^H}{2} \left[ 1 + \sqrt{\frac{2}{\pi H (1-d)}} \right],
\end{equation}
where in the last passage we used the Stirling approximation being $H(1-d)\gg1$.
Then, we have $\tilde{n}(H(1-d)/2,d) \gtrsim 1/2$, and similar calculations can be drawn for smaller thresholds, e.g.,  $\tilde{n}(H(1-d)/2 -1 ,d) \lesssim 1/2$.

\begin{figure}[tb] \begin{center}
\includegraphics[width=.5\textwidth]{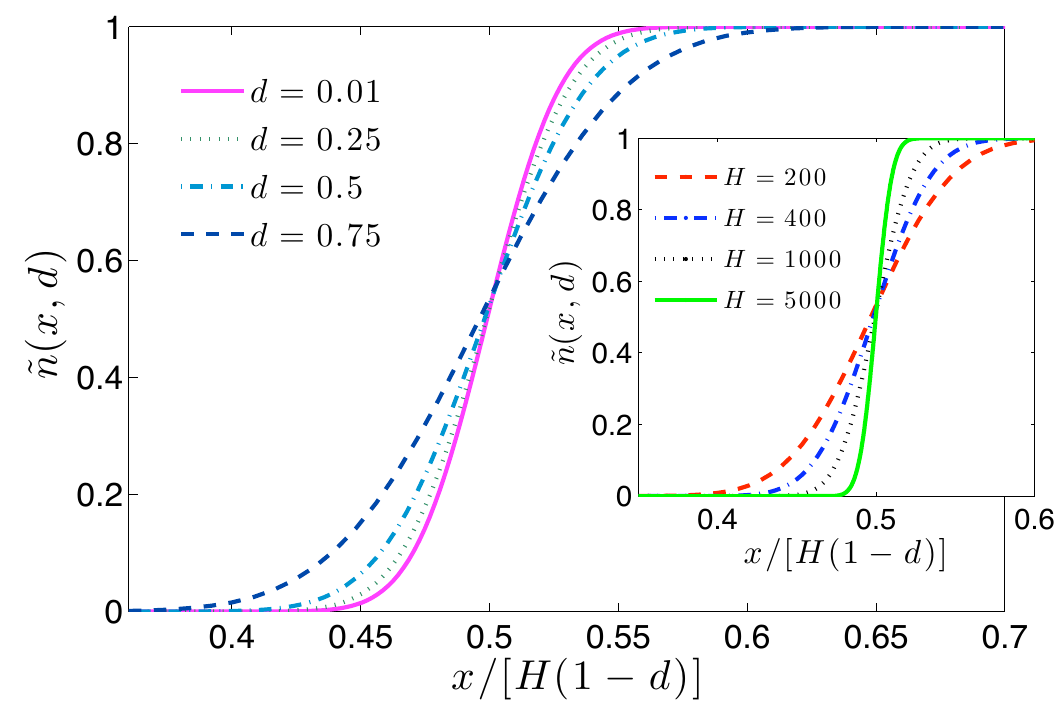}
\caption{\label{fig:percola} (Color on line) Normalized number of accessible configurations $\tilde{n}(x,d)$ as a function of $x$ and $d$ for a system made up of $H=$ lymphocytes. The critical line $x_c=(1-d)$, corresponds to the emergence of a giant component and above it the system is no longer able to retrieve patterns of information.}
\end{center}
\end{figure}

As shown in Fig.~\ref{fig:percola}, once $d$ is fixed, when $x$ is small only a microscopic fraction $\tilde{n}(x,d)/2^H$ of configurations is accessible (in the thermodynamic limit this fraction is vanishing), while by increasing the tolerance $x$, more and more configurations get accessible and correspondingly their fraction gets macroscopic. From a different perspective, each configuration can be looked at as a node of a graph and those accessible are connected together. The link probability is then related to $x$ and when $x$ is large enough a ``giant component'' made up of all accessible configurations emerges.
This is a percolation process in the space of configurations.
Indeed, similarly to what happens in canonical percolation processes, the curves representing the giant component relevant to different sizes $H$ intersect at around $1/2$, and this determines the percolation threshold $x_c$.
According to Eq. (\ref{eq:threshold}) we can write $x_c\approx H(1-d)/2$.

%

Interestingly, when a giant component emerges, retrieval is no longer meaningful because the system may retrieve essentially anything and this corresponds to the critical line (in the $d,\beta$ plane) where all the magnetization simultaneously disappear.

\section{Perturbing with random fields: Lymphocytosis and autoimmunity}

The model described so far is able to capture several issues of real immune systems. For instance, when affected by Autoimmune Lymphoproliferative Syndrome, the system undergoes a massive activation where a fraction (possibly extensive) of clones are made to expand (e.g., see \cite{alps}); in this particular condition the reference state has to be taken $b_0 \gg 0$ to mimic strong clonal expansions.  Therefore, in the partition function $Z_{H,B}(\alpha,\beta)$, we can substitute the centered Gaussian weight $\exp(-\sum_{\mu}b_{\mu}^2/2)$ with $\exp[-\sum_{\mu}(b_{\mu}-b_0)^2/2]$ for the pertaining fraction of deranged clones. It is immediate to check \cite{JTB} that when the bipartite spin-glass system is mapped into the associative network counterpart, this time a new term appears in the resulting Hamiltonian $\tilde{\mathcal{H}}(h;\xi)$:
\begin{equation}\label{toy1}
\tilde{\mathcal{H}}(h;\xi) = - \frac{1}{2H} \sum_{i,j=1}^H \left(\sum_{\mu=1}^B \xi_i^{\mu}\xi_j^{\mu} \right)h_i h_j - a \sum_{i=1}^H \eta_i h_i,
\end{equation}
where $\eta \in \mathcal{N}[0,1]$ and $a$ is a parameter tuning the overall effect (it includes the number of deranged clones and their size $b_0$).

It is intuitive to see that, as the parameters ($\beta$, $a$) are tuned, different behaviors emerge. For instance, when $a$ is too large one expects that random effects prevail over the retrieval capacity of the system and helpers are no longer able to properly manage an immune response, ultimately leading to random activation of B-clones and possible autoimmunity phenomena.
Indeed, it has been recently evidenced (see e.g. \cite{limpo1,limpo2}) that lymphocytosis can yield Chronic Lymphocytic Leukemia, which in turn is often
accompanied with mild-to-severe autoimmune manifestations \cite{limpo3,limpo4}. However,
a clear explanation for this link is still missing. Hence, in the following, we try to obtain a systemic rationale of the relation between lymphocytosis and autoimmunity through the statistical mechanics perspective.

Given the Hamiltonian (\ref{toy1}), the goal is finding an explicit expression for the self-consistencies of the order parameters $\langle m_{\mu}\rangle_{\xi}$ that generalizes Eq. $(27)$.
To accomplish this task, we adapted the Hamilton-Jacobi method, originally developed in the framework of spin glasses \cite{barraHJ2,barraHJ1}
to this scenario: Let us consider a generalized partition function $Z_{H,B}(t,x)$ which depends on two
interpolating parameters $t, x$, that can be considered as \emph{generalized time} and \emph{space}, such that the corresponding
averaged free energy $\langle F(t,x) \rangle_{\xi, \eta}$ can be derived from the free energy at finite $H$ as $\langle F(t,x) \rangle_{\xi, \eta} =\lim_{H \to\infty}
\langle F_H(t,x)\rangle_{\xi, \eta}$, being
\begin{equation}
\langle F_H(t,x)\rangle_{\xi, \eta} = \frac{-1}{H}\mathbb{E}\ln{Z_{H,B}(t,x)}= -\frac{1}{H}\mathbb{E}\ln\sum_{\{h\}}e^{\frac{t}{2H}\sum_{i,j=1}^H\sum_{\mu=1}^B\xi_i^{\mu}\xi_j^{\mu}h_i h_j + x \sum_{i=1}^H h_i +\beta a \sum_{i=1}^H\eta_i h_i}.
\end{equation}
Note that the correct free energy is recovered when $t=\beta,x=0$. It is straightforward to consider explicitly the $\langle F_H(t,x)\rangle_{\xi, \eta}$ derivatives
\begin{align}
&\frac{\partial \langle F_H(t,x)\rangle_{\xi, \eta}}{\partial t}=-\frac{1}{2}\sum_{\mu=1}^B \langle m_{\mu}^2 \rangle_{\xi,\eta},\\
&\frac{\partial \langle F_H(t,x)\rangle_{\xi, \eta}}{\partial x_{\mu}}= \langle m_{\mu} \rangle_{\xi,\eta},
\end{align}
and note that, if we define a \emph{potential} $V_H(t,x)$ as the sum of the variances of all the $m_{\mu}$, namely
\begin{equation}
V_H(t,x)=\frac{1}{2}\sum_{\mu}^B \left( \langle m_{\mu}^2 \rangle_{\xi,\eta}-\langle m_{\mu} \rangle_{\xi,\eta}^2 \right),
\end{equation}
in the space of the interpolants $(t,x)$,  the following Hamilton-Jacobi equation holds
\begin{equation}
\frac{\partial \langle F_H(t,x)\rangle_{\xi, \eta}}{\partial t}+ \frac{1}{2}\sum_{\mu=1}^B \bigg(\frac{\partial \langle F_H(t,x)\rangle_{\xi, \eta}}{\partial x_{\mu}}\bigg)^2 +V_H(t,x)=0.
\end{equation}
Then, solving the model consists in  finding the free-field solution, requiring $V_H(x,t)=0$  as in the thermodynamic limit ($H \to \infty$) the order parameters self-average.
If the potential is zero then the energy is a constant of motion and it is trivially the \emph{Lagrangian}
$\mathcal{L} = \frac{1}{2}\sum_{\mu}^B(\frac{\partial \langle F_H(t,x)\rangle_{\xi, \eta}}{\partial x_{\mu}})^2$. Further, the trajectories of motion are straight lines
\begin{equation}
x_{\mu}(t) =x_0 + \langle m_{\mu}\rangle_{\xi,\eta} t.
\label{eq:retta}
\end{equation}
If we denote with a bar the Hamilton function which satisfies the free-field problem, such solution $\bar{F}(t,x)$ can
 be worked out evaluating it in a starting point $t_0,x_0$ in the $(t,x)$-space and adding to it the integral of the Lagrangian over the time, namely
\begin{equation}
\langle \bar{F}(t,x) \rangle_{\xi, \eta} =\langle \bar{F}(t_0,x_0)\rangle_{\xi, \eta} + \int_{t_0}^t dt'\mathcal{L}(t',x).
\label{eq:hamiltonsol}
\end{equation}
We choose $t_0=0$ and we have
\begin{eqnarray}
\nonumber
\langle \bar{F}(0,x_0)\rangle_{\xi, \eta} &=& -\frac{1}{H}\mathbb{E}\ln\sum_{\{\sigma\}}\exp{\left( \sum_{\mu=1}^B x_0^{\mu}\sum_{i=1}^H\xi_i^{\mu}\sigma_i +\beta a\sum_{i=1}^H\eta_i\sigma_i \right) } \\
&=&-\frac{1}{H}\mathbb{E}\ln \prod_{i=1}^H\sum_{\{\sigma\}}\exp \left(\sum_{\mu=1}^B x_0^{\mu}\xi_i^{\mu}\sigma_i +\beta a \eta_i\sigma_i \right).
\end{eqnarray}
Using equation (\ref{eq:retta}) we obtain
\begin{equation}
\langle \bar{F}(0,x_0)\rangle_{\xi, \eta}=-\ln{\left \{Ê2\cosh\left[{\sum_{\mu=1}^B(x(t)- \langle m^{\mu}\rangle_{\xi,\eta} t)\xi_i^{\mu} +\beta a \eta}\right] \right\} }.
\end{equation}
For the second term of Eq. (\ref{eq:hamiltonsol}),  since $V_H(t,x)=0$ when $H\to\infty$, the Lagrangian takes the ''standard" form $\mathcal{L}= p^2 /2m$, where the mass is $m=1$ and the squared momentum $p^2=\sum_{\mu}^B \langle m_{\mu}^2 \rangle_{\xi,\eta}$. Thus, overall we can write
\begin{equation}
\int{dt' \mathcal{L}(t')}= \sum_{\mu}^B \langle m^{\mu}\rangle_{\xi,\eta}^2 \frac{t}{2}.
\end{equation}
Now, we must evaluate the solution at $t=\beta, x=0$:
\begin{equation}
\langle \bar{F}(\beta,d,a) \rangle_{\xi, \eta} =\ln2+\ln\cosh(\beta\sum_{\mu} \langle m^{\mu}\rangle_{\xi,\eta} \xi_i^{\mu} +\beta a \eta) - \frac{\beta}{2}\sum_{\mu}^B \langle m^{\mu} \rangle^2_{\xi,\eta}.
\end{equation}
So $\langle \bar{F}(\beta,d,a)\rangle$ corresponds to the free energy
of the system perturbed by the random field $\eta$ and by minimizing this function  with respect
to $\langle m_{\mu}\rangle_{\xi,\eta}$,  hence posing
$$\frac{\partial\langle \bar{F}(\beta,d,a)\rangle_{\xi, \eta}}{\partial \langle m^{\mu}\rangle_{\xi,\eta}}=0$$
we find the self-consistency equations
\begin{equation}\label{randomM}
\langle m^{\mu}\rangle_{\xi,\eta}= \langle\xi^{\mu}\tanh{\beta(\sum_{\mu} m^{\mu}\xi_i^{\mu}+a\eta)}\rangle_{\xi,\eta},
\end{equation}
which generalize Eq. $(27)$ and recover the latter when  $a \to 0$.
Finally, we average $\langle m^{\mu}\rangle_{\xi,\eta}$ over $P(\xi)$ and $P(\eta)$ and solve the equations numerically, as we are going to show explicitly for $B=2$.

\begin{figure}[tb] \begin{center}
\includegraphics[width=0.6\textwidth]{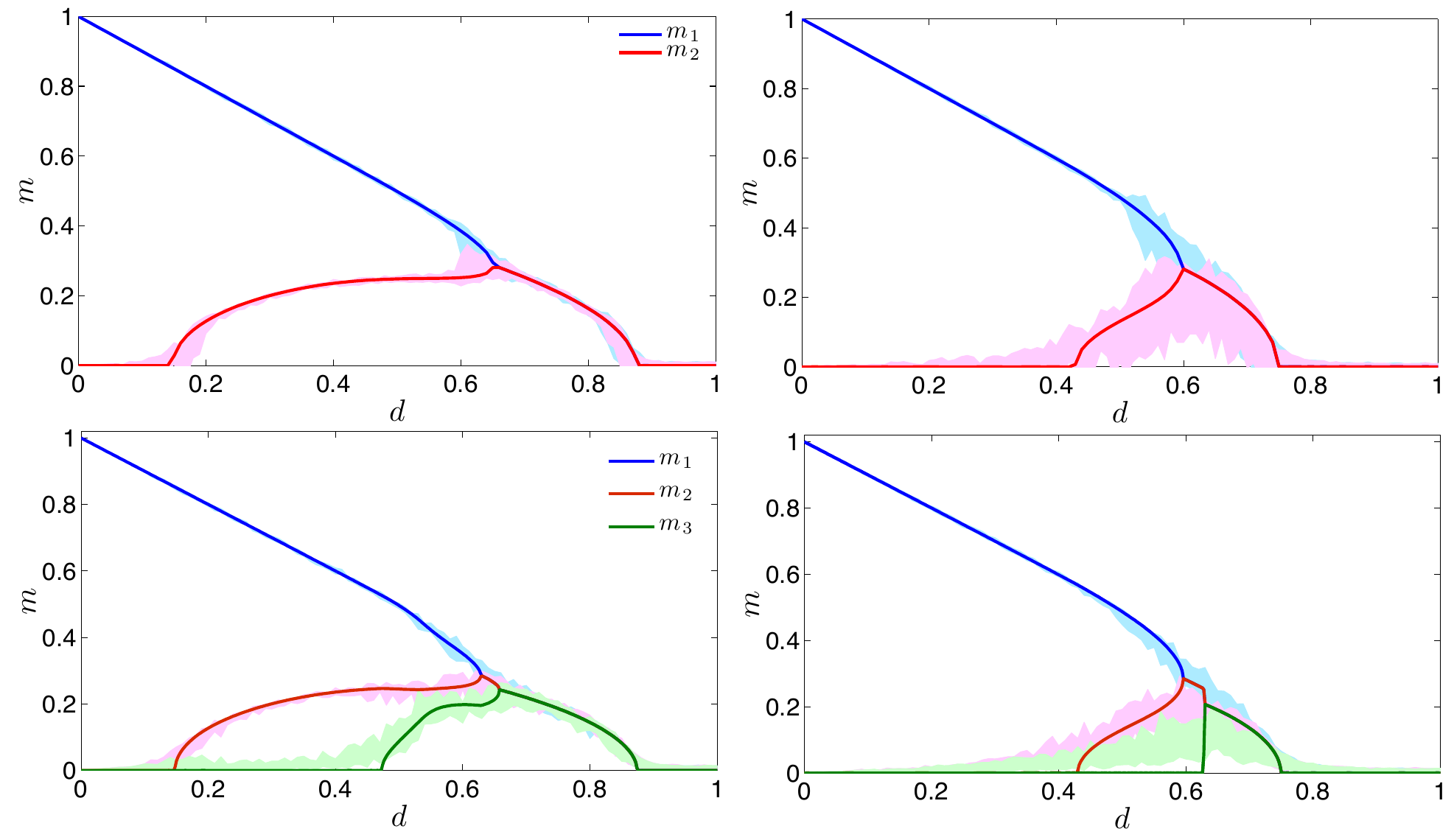}
\caption{\label{fig:silvia1} (Color on line) Comparison of the parallel processing capabilities of the network with two (upper panels) and three (lower panels) retrieved patterns and a random field afflicting their recognition. Left panels shows the order parameter $\textbf{m}$ versus dilution at noise level $\beta=10$ and $a=0.1$. Right panels shows the order parameter $\textbf{m}$ versus dilution at noise level $\beta=100$ and $a=0.2$. Both analytical (solid lines) and Monte Carlo (the shades represent the standard deviation) results are displayed to show the good overlap between the various curves. The net effect of the random field is always to deteriorate the net capabilities of helper network.}
\end{center}
\end{figure}

Solving Eq. (\ref{randomM}) in the case of two patterns we find
\begin{align}
\langle m_1 \rangle_{\xi,\eta} = &\int{d\mu}(\eta)\frac{d(1-d)}{2}\left \{ \tanh [\beta(\langle m_1 \rangle_{\xi,\eta} +a\eta)] +\tanh[\beta(\langle m_1 \rangle_{\xi,\eta} -a\eta)]\right\} +
\\&+\frac{d^2}{4}\left \{ \tanh[\beta(\langle m_1 \rangle_{\xi,\eta} + \langle m_2 \rangle_{\xi,\eta} +a\eta)] + \tanh[ \beta(\langle m_1 \rangle_{\xi,\eta} + \langle m_2 \rangle_{\xi,\eta} -a\eta)] \right \}+
\nonumber \\
&+\frac{d^2}{4}\left \{ \tanh[\beta(\langle m_1 \rangle_{\xi,\eta} - \langle m_2 \rangle_{\xi,\eta} +a\eta) ] + \tanh[ \beta(\langle m_1 \rangle_{\xi,\eta} - \langle m_2 \rangle_{\xi,\eta} -a\eta)] \right \}, \nonumber
\end{align}
\begin{align}
\langle m_2 \rangle_{\xi,\eta} = &\int{d\mu}(\eta)\frac{d(1-d)}{2}\left \{ \tanh[\beta(\langle m_2\rangle_{\xi,\eta} +a\eta)]+\tanh[\beta(\langle m_2\rangle_{\xi,\eta} -a\eta)] \right \}+
\\&+\frac{d^2}{4}\left \{ \tanh[ \beta(\langle m_1\rangle_{\xi,\eta} + \langle m_2\rangle_{\xi,\eta} +a\eta)] + \tanh[ \beta(\langle m_1 \rangle_{\xi,\eta} + \langle m_2 \rangle_{\xi,\eta} -a\eta)] \right \}+\nonumber \\
&-\frac{d^2}{4}\left \{ \tanh[\beta(\langle m_1 \rangle_{\xi,\eta} - \langle m_2 \rangle_{\xi,\eta} +a\eta)] +\tanh[\beta(\langle m_1 \rangle_{\xi,\eta} - \langle m_2 \rangle_{\xi,\eta} -a\eta)] \right\}.\nonumber
\end{align}

Looking at Eq. (\ref{randomM}), it is easy to see that when the value of $a$ prevails on the retrieval counterpart, helpers essentially behave randomly, giving wrong instructions to B-clones, ultimately implying autoimmune manifestations. Hence, autoimmunity and lymphocytosis appear as deeply correlated phenomena.

In Fig.~\ref{fig:silvia1} we show how the overlaps depend on $d$ in the case $B=2$ and $B=3$ and for different values of noise and field. We also successfully compare the numerical solution of self-consistence equations with results from numerical simulations.

\section{Summary and outlooks}

In a recent paper we proposed a model for the adaptive immune response, where helpers and B-cells interact via cytokines and are described as a fully-connected bipartite spin glass; we also showed that such a model is equivalent to an attractor associative network where helpers are able to collectively orchestrate the activation of B-cells \cite{JTB}. This  network, although able to capture several issues of real immune systems, was actually able to elaborate only one strategy at a time, namely, helpers could manage each clonal lineage of B-cells sequentially.

Here we extended the model by introducing a degree of dilution $d$ in the bipartite spin-glass, in such a way that only a fraction of the whole B-repertoire interacts with a given helper lineage; this yields a much more biological description, and gives rise to a remarkable emergent behavior.
In particular, we show that this system is able to arrange multiple strategies simultaneously, namely, helpers are able to orchestrate and coordinate the responses of several B-clones at the same time. This is very consistent with the well-known capability of the immune system to contemporary fight several pathogens.

We studied in detail the case where the amount $B$ of B-clones is sub-linear with respect to the amount $H$ of helpers, namely $\lim_{H \to \infty}(B/H)=\alpha=0$. This is certainly an oversimplification, yet the novelty and the potentiality of this approach are already evident and this may contribute to a rationale understanding of the systemic properties of lymphocyte networks.

From a technical point of view, we studied the model via statistical mechanics solving for the free-energy and obtaining, through its extremization, the self-consistencies for the order parameter.
These equations have been hierarchically solved and tested against the results obtained via signal-to-noise analysis and Monte Carlo simulations, finding overall perfect agreement.

Once showed that the ``pure state ansatz" of standard associative networks can no longer minimize the free energy, we introduced a ``parallel ansatz",  which works at relatively small degrees of dilution and we studied the stability of the basin of attraction of the minima it generates.

Finally, we investigated  the case of strong clonal expansion (lymphocytosis) which results in adding an extra random-term to the system and we solved this generalization of the model through the adaptation of the Hamilton-Jacobi technique. Again, we checked results against Monte Carlo simulations finding excellent agreement. The biological interpretation of these findings suggests that there is a deep, systemic, link between lymphocytosis and autoimmunity, consistently with clinical and experimental evidence.

Future works, beyond the microscopical interpretation of the tunable parameters, should be focused on the saturated case, namely $\lim_{H \to \infty}(B/H)=\alpha >0$, which is still mathematically challenging. The presence of antigens (fields) and a discrimination between B-clones with low/high avidity against self tissues are also in order to show further emerging properties concerning self/non-self discrimination.

\section*{Acknowledgments}

The authors are grateful to Guido Valesini and Rossana Scrivo for useful conversations. \\
This research was sponsored by the FIRB grant RBFR08EKEV.\\
Sapienza Universit$\grave{a}$ di Roma and Istituto Nazionale di Fisica Nucleare are acknowledged too for partially supporting the work.

\section*{Appendix: Numerics}

In this Appendix we discuss details on Monte Carlo simulations.
\newline
All the simulations were performed on a system Ubuntu Linux with Intel Core I7, $3.2$Ghz, $12$ CPU, Nvidia-Fermi technology, $12$ Gb RAM and OpenMP libraries.
The simulations were carried out sequentially according to the following algorithm:
\begin{enumerate}
\item Building and storaging of the coupling matrix.

First, we generate $B$ patterns according to the distribution ($d=0$):
\be
P(\xi_i^{\mu})=\frac{1}{2}\delta_{(\xi_i^{\mu}-1)}+\frac{1}{2}\delta_{(\xi_i^{\mu}+1)},
\ee
then, we build a char-matrix $J_{ij} = \sum_{\mu} \xi_i^{\mu} \xi_j^{\mu}$ with entries ranging $\in [0,2B+1]$ and acting as key pointing to another hash-matrix $\tilde{J}_{ij}$ where the $H(H-1)/2$ real numbers accounting for the Hebb interactions (see Eq. $(16)$) are stored. If the amount of patterns do not exceed $B=256$, i.e. one byte, it is then possible to account for $10^{5}$ helpers with no need of swapping on hard disk (which would sensibly affect the performance of the simulation). This condition is fulfilled for the low storage regime we are interested in.

\item Initialize the network status.

We checked the two standard approaches: The first is to initialize the network in a (assumed) fixed point of the dynamics, namely
\be
h_i=\xi_i^{1} \,\,\,\,\forall i \in [1,...,H],
\ee
and check its evolution: This gives information on the structure of the basins of attraction of the minima as we vary the dilution (see Point $5$).
\newline

The second approach is to initialize the network randomly: We set $h_i = 1$ with probability $0.5$ and $h_i=-1$ otherwise. This is a standard procedure to follow the relaxation to a fixed point with no initial assumption and gives information on the structure of the basins of attraction of the minima at fixed dilution.

\item Evolution dynamics

The activity of helpers evolves according to a standard (random and sequential) Glauber dynamics for Ising-like systems \cite{amit}:
At each time interval, the state of a lymphocyte  is updated according to its input signals, where the probability of the unit's activity is equal to a rectified value of the input (logit transfer function), i.e.
\begin{equation}
Pr(h_i(t)=\pm1)=\frac{1}{1+\exp[\mp 2\beta \sum_j J_{ij}h_j]}.
\end{equation}
The field-updating process is managed by a linked list whose parsing is parallelized through OpenMP.

\item Convergence of the simulation.

Due to the peculiar structure of the fields induced by pattern dilution (see Fig.~$3$, right panel), the field insisting on a given helper may be zero and the related spin would flip indefinitely. To avoid this pathological situation we skip the updating of these "paramagnetic" lymphocytes and focus on the remaining ones: In the zero noise limit convergence is almost immediate, such that when the whole ensemble of helpers remains unchanged for the whole $N$-length of the update cycle, dynamics is stopped and the resulting  B pattern overlaps are printed on a file.

Relaxation at non-zero noise is checked through the linked list (see next step):  The pointer of each helper that is aligned with its own field is stored, the ones of helpers with no net fields are removed from the linked list, while all the other helpers mismatched to their own fields, are added into the linked list.

\item Making the $B$ patterns sparser.

There can be two deeply different ways of increasing dilution. The former is a Bernoullian approach and essentially if one starts from a dilution $d = 0.45$ toward a dilution $d=0.5$ essentially may forget the starting information and generate a random pattern with on average one half of zero entries; the latter is a Markovian dilution by which one needs to start from the previous coupling matrix (and patterns) diluted at $d=0.45$ and increases dilution on that structure.
%
\newline
Dilution is tuned at steps of $0.01$, ranging from $d=0$ to $d=1$.
\newline
We take as the state of the network the last equilibrium state, then go to point (3).
\end{enumerate}
Through Markovian dilution, we can follow the evolution of the pure Hopfield attractors
while tuning $d$. In general, the results obtained via numerical simulations are in perfect agreement with the theory.

\section*{Appendix: Signal to noise ratio in the zero fast noise limit}

As usually done in the neural network context \cite{amit}, we couple the statistical mechanics inspection to signal-to-noise analysis. Aim of this procedure is trying to confirm the ``parallel ansatz" we  made by studying the stability of the basins of attractions (whose fixed points are the learnt strategies) created in the hierarchical fashion we prescribed. We recall that the model we are investigating describes a low storage of information in the associative network so that no slow noise is induced by the underlying spin glass, i.e. $\alpha = 0$. Nonetheless, we study the signal to noise ratio in the zero fast noise limit  ($\beta \to \infty$) as a problem formulated in general terms of $\alpha, d$; then, we take the limit $\alpha \to 0$ to get an estimate about the stability of the basins of attractions (where the presence of fast noise can possibly produce fluctuations).

Without loss of generality, we assume that the best retrieved pattern is the first one. This means that spins are aligned with the non-null entries in the first bit-string $\xi^1$, while the remaining spins explore the  other patterns. Thus, for the generic spin $h_i$ we can write
\be
h_i=\xi_i^1+\sum_{\nu=2}^B \xi_i^{\nu}\prod_{\mu=1}^{\nu-1}\delta(\xi_i^{\mu}).
\ee
Accordingly, the local field acting on the $i^{th}$ lymphocyte can be written as
\be
\varphi_i=\frac{1}{H}\sum_{j \neq i}^{H}\sum_{\mu=1}^{B} \xi_i^{\mu}\xi_j^{\mu}[\xi_j^1+\sum_{\nu=2}^B \xi_j^{\nu}\prod_{\mu=1}^{\nu-1}\delta(\xi_j^{\mu})].
\ee
\begin{itemize}

\item In the reference case $B=1$, similarly to the pure states of the Hopfield network, we set
\be
h_i=\xi_i^1+\delta(\xi_i^1)k_i,
\ee
where $k_i$ is a random variable uniformly distributed on the values $\pm 1$
added to ensure that there are no nulls entries in the state of the network. Hence we find
\be
\langle \varphi_i h_i\rangle_{\mathbf{\xi}}=\langle signal + noise  \rangle_{\mathbf{\xi}}=\langle signal\rangle_{\mathbf{\xi}},
\ee
being $\langle noises\rangle_{\mathbf{\xi}}=0$, and so for large $H$ we have
\be
\langle signal \rangle_{\mathbf{\xi}}=\frac{H-1}{H}(1-d)=(1-d),
\ee
while
\be
\langle (noises)^2\rangle_{\mathbf{\xi}}=\frac{B-1}{H}(1-d)^2=\alpha (1-d)^2.
\ee
\item In the test case of two patterns retrieved, $B=2$, we set:
\be
h_i=\xi_i^1+\delta(\xi_i^1)[\xi_i^2+\delta(\xi_i^2)k_i].
\ee
Now, we need to distinguish between the various possible configurations:
\begin{itemize}
\item $\forall i$ such that $\xi_i^1\neq 0 ,\xi_i^2 =0 $ and so that $h_i=\xi_i^1 \neq 0$
for large value of $H$
\be
\langle signal \rangle_{\mathbf{\xi}}=(1-d), \ \ \ \ \langle noises\rangle_{\mathbf{\xi}}=0,
\ee
\be
\langle (noises)^2\rangle_{\mathbf{\xi}}=\frac{(H-1)(B-2)}{H^2}(1-d)^2=\alpha (1-d)^2.
\ee
\item $\forall i$ such that $\xi_i^1\neq 0 ,\xi_i^2 \neq 0 $ and so that $h_i=\xi_i^1 \neq 0$
\newline
if $\xi_i^1=\xi_i^2$
\be
\langle signal \rangle_{\mathbf{\xi}}=2(1-d)-(1-d)^2, \ \ \ \ \langle noises\rangle_{\mathbf{\xi}}=0,
\ee
if $\xi_i^1=-\xi_i^2$
\be
\langle signal \rangle_{\mathbf{\xi}}=(1-d)^2, \ \ \ \ \langle noises\rangle_{\mathbf{\xi}}=0.
\ee
and in both cases
$$
\langle (noises)^2\rangle_{\mathbf{\xi}}=\frac{(H-1)(B-1)}{H^2}(1-d)^3+\frac{(H-1)(B-2)}{H^2}d(1-d)^2=\alpha (1-d)^2.
$$

\item  $\forall i$ such that $\xi_i^1= 0 ,\xi_i^2 \neq 0 $ and so that $h_i=\xi_i^2 \neq 0$
\be
\langle signal \rangle_{\mathbf{\xi}}=d(d-1), \ \ \ \ \langle noises\rangle_{\mathbf{\xi}}=0,
\ee
\be
\langle (noises)^2\rangle_{\mathbf{\xi}}=\frac{(H-1)(B-1)}{H^2}(1-d)^3+\frac{(H-1)(B-2)}{H^2}(1-d)^2d=\alpha (1-d)^2.
\ee
\end{itemize}
\end{itemize}
Therefore, in the regime of low storage of strategies considered ($\alpha=0$), the retrieval is stable, states are well defined and the amplitude of the signal on the first channel is order $(1-d)$ while on the second is of order $d(1-d)$, in perfect agreement with both the statistical mechanics analysis and Monte Carlo simulations.


\end{document}